\documentclass[preprint,showkeys,secnumarabic,amsfonts,showpacs,amsmath,amssymb]{revtex4}
\usepackage[dvips]{color}
\usepackage{array}
\usepackage{rotating}
\usepackage{epsfig}
\usepackage{amsmath}
\usepackage{graphicx}
\usepackage{graphics}

\begin{document}
\title{Interacting Tsallis agegraphic dark energy in DGP Braneworld Cosmology}

\author{Zahra Feizi Mangoudehi}
\email{zahraf@msc.guilan.ac.ir} \affiliation{Department of Physics, University of Guilan, Rasht, Iran}
\date{\today}

\begin{abstract}
 \noindent \hspace{0.35cm}
 The purpose of this paper is to study the Tsallis agegraphic dark energy with an interaction term between dark energy and dark matter in the DGP brane-world scenario. For this, we assume some initial conditions to obtain the dark energy density, deceleration, dark energy EoS, and total EoS parameters. Then, we analyze the statefinder parameters, $\omega'{}_{DE}-\omega_{DE}$ plots, and classical stability features of the model. The results state that the deceleration parameter provides the phase transition from decelerated to accelerated phase. The $\omega_{DE}$ graphs show the phantom behavior, while the $\omega_{tot}$ exhibits the quintessence and phantom during the evolution of the Universe. Following the graphs, the Statefinder analysis shows the quintessence behavior of the model for the past and present. However, it tends to the $\Lambda CDM$ in the following era. The $\omega'{}_{DE}-\omega_{DE}$ plot indicates the thawing or freezing area depending on the type of era and different values of $b^{2}$, $\delta$, and $m$. By the square of the sound speed, we see the model is stable in the past, stable or unstable at the current time, and unstable in the future for selected values of $b^{2}$, $\delta$, and $m$. To test the model, we use the recent Hubble data. We also employ Akaike Information Criterion (AIC) and Bayesian Information Criterion (BIC) to compare the model with the $\Lambda CDM$ as the reference model. In addition, we test the model using the $H-z$ plot, and we see a turning point in the future time. The results from the best fit values for the $\omega_{tot}$ plot emphasize that the Universe is in the quintessence region in the current time. It will enter the phantom phase, and then it will approach the $\Lambda$ state in the future. But, the $\omega_{DE}$ always stays on the phantom region. The model is unstable in the present and progressive era.
\end{abstract}

\pacs{04.50.Kd; 98.80.-k}

\keywords{TADE, Interaction, DGP braneworld cosmology, Statefinder diagnostic, The $\omega^{'}_{DE}-\omega_{DE}$ plane, Observational constrains, Turning point}
\maketitle

\section{Introduction}
\label{intro}
Nowadays, the accelerated expansion of the universe has been confirmed by various types of observational data, especially supernovae type Ia \cite{bib1,bib2,bib3,bib4,bib5}. To explain such a phase of acceleration, two chief approaches are considered: introducing a new unknown cosmological ingredient called $dark$ $energy$ (DE) or modifying the gravitational part of the Einstein equations \cite{bib6,bib7,bib8}.

However, the most well-accepted description of the current universe is the use of the dark energy idea. As we know, dark energy can count as responsible for accelerating the universe due to having a negative pressure. So, Although the nature of dark energy is ambiguous, some candidates can be regarded for its negative pressure feature.

The simplest candidate for dark energy is called the cosmological constant \cite{bib9,bib10}, which is a key component in the $\Lambda CDM$ model, and it identifies with $p=-\rho$. Although the $\Lambda CDM$ model is consistent with observational data, it can lead to some cosmological problems, such as coincidence problem (\cite*{bib11}). So, a series of alternative theoretical DE models have been proposed (for example \cite{bib12,bib13,bib14}). But any new model of DE has many hidden features and can create new problems by introducing new degrees (s) of freedom.

In terms of introducing dark energy, \cite{bib15} proposed the agegraphic dark energy (ADE) model by considering the Universe age (T) as IR cutoff. This model is based on the line of quantum fluctuations of spacetime and the well-known time-energy uncertainty relation. Then, \cite{bib16} proposed a new model of agegraphic dark energy (NADE) based on the uncertainty connection of quantum mechanics with considering the conformal time ($\eta$) as the time scale of the FRW Universe. \cite*{bib17} studied the new agegraphic dark energy (NADE) model by using the observational data of supernovae type Ia, the cosmic microwave background (CMB), and the baryon acoustic oscillation (BAO). \cite{bib18} investigated the ADE and NADE models in the framework of Brans-Dicke theory. The authors have derived the dark energy equation of state ($w_{DE}$) and the deceleration parameter ($q$) in their models. \cite*{bib19} explored the interacting and non-interacting NADE scenario in a Universe governed by Horava-Lifshitz cosmology. They have extracted the differential equation that determines the evolution of the DE density parameter. \cite{bib20} implemented agegraphic and new agegraphic dark energy models with and without interaction terms between dark matter and dark energy in a cyclic Universe. The author has found that the ADE and NADE models without interaction terms can not provide a cyclic Universe. Then, he has shown that the interacting new agegraphic dark energy model can produce a cyclic Universe better than the interacting agegraphic dark energy model. Later, two authors introduced a new black hole thermodynamical entropy which is called $Tsallis$ $entropy$ (\cite{bib21}). This entropy is inspired by quantum gravity considerations leading to a fractal structure of horizon (\cite{bib22}). Indeed, the Tsallis statistics is beneficial for fractal systems, and Barrow finding is a quantum gravitational sign for a relation between quantum features of gravity and non-extensive statistics \cite{bib23,bib24}. Then, \cite{bib25} proposed a new dark energy model using the non-extensive Tsallis entropy and the holographic hypothesis. They have introduced Tsallis agegraphic dark energy (TADE) with time scale as infrared IR cutoff. Also, the authors have proposed the New Tsallis agegraphic dark energy (NTADE) model by considering conformal time $\eta$ as the IR cutoff. In this work, the authors have studied the TADE and NTADE models with and without interaction terms for investigating the behavior of $q$, $w_{DE}$, $\Omega_{DE}$ (dark energy density parameter), and $v_{s}^2$ (the squared of sound speed) during the cosmic evolution. \cite{bib26} discussed the Tsallis agegraphic dark energy in terms of studying the evolution of statefinder parameters and $w^{'}_{DE}-w_{DE}$ planes. \cite{bib27} explored the TADE model by assuming a sign-changeable interaction between Tsallis agegraphic dark energy and dark matter. On the ground, the author has obtained the deceleration ($q$), dark energy density ($\Omega_{DE}$), dark energy EoS ($w_{DE}$), and total EoS ($w_{tot}$) parameters. Also, he has analyzed the statefinder pair and the $w_{DE},w^{'}_{DE}$ plot.

Moreover, various modified theories of gravity have expanded to explain the accelerated expansion of the universe.
Braneworld scenario is an interesting approach to modify the Einstein theory. This scenario included Dvali-Gabadadze-Porrati (DGP) braneworld, Randall and Sundrum (RS II), and the Cyclic models. Among these branches of the braneworld cosmology, the Dvali-Gabadadze-Porrati (DGP) braneworld model \cite{bib28,bib29,bib30} counts as the most favorable model. This model is a five-dimensional bulk with self-accelerating and normal characteristics. The self-accelerating characteristic of the DGP model is able to show the late-time acceleration \cite{bib31,bib32,bib33,bib34,bib35,bib36,bib37,bib38,bib39}. But, this characteristic of the DGP model cannot satisfy the phantom divide crossing by itself. On the other hand, the normal DGP branch without a dark energy component cannot describe the acceleration phase, but it can show the phantom-like phase on the brane. Thus, to solve this problem, employing dark energy (\cite{bib35}) is essential.

Motivated by the above study, we would like to investigate the interacting Tsallis agegraphic dark energy in the DGP braneworld cosmology. We are interested in exploring the evolution of dark energy density, equation of state, and deceleration parameters against the redshift. Then, we will determine the statefinder diagnostic parameters, $w_{DE},w^{'}_{DE}$ diagram, and classical stability for our model. For this aim, we have assumed $\Omega_{DE}(z=0)=0.680$ (TT+lowE data), $H(z=0)=66.98$ (TT+lowE data) \cite{bib40}, and $\Omega_{r_{c}}=0.0003$ (\cite*{bib41}). Other values are considered changeable. Furthermore, the complete analysis of the model using the $\chi^{2}_{min}$ method of recent Hubble data (\cite*{bib42}) will be employed.

The present work is structured as follows: In Sect. 2, we have introduced the interacting Tsallis agegraphic dark energy in the DGP braneworld scenario. In Sect. 3, we have studied the cosmological parameters, such as the deceleration and EoS parameters. Then, in Sects. 4, 5, and 6, we have discussed the Statefinder, $w^{'}_{DE}-w_{DE}$ plane, and stability of the model. Finally, in Sect. 7, we have fitted our model
with the Hubble observational data.

\section{THE ITADE IN DGP MODEL}
\label{sec:1}
A homogeneous and isotropic Friedmann-Robertson-Walker universe can be
described by
\begin{eqnarray}
ds^{2}=-dt^{2}+a^2(t)(\frac{dr^{2}}{1-Kr^{2}}+r^{2}d\Omega^{2}),
\end{eqnarray}
in which $K$ shows a flat, closed, or open universe with 0, 1, and -1 values, respectively. The modified Friedmann equation in the DGP braneworld model is given by \cite{bib43,bib44,bib45,bib46,bib47}
\begin{eqnarray}
H^{2}+\frac{K}{a^{2}}=(\sqrt{\frac{\rho}{3M_{pl}^{2}}+\frac{1}{4r_{c}^{2}}}+\frac{\epsilon}{2r_{c}})^{2},
\end{eqnarray}
where $\rho$ is the total energy density included $\rho_{m}$ (dark matter density) and $\rho_{DE}$ (dark energy density) with the following conservation equations
\begin{eqnarray}
\dot{\rho}_{m}+3H(\rho_{m}+p_{m})=Q,~~~~~~~
\\
\dot{\rho}_{DE}+3H(\rho_{DE}+p_{DE})=-Q.&&~~~~~~~~~~~~~~~
\end{eqnarray}

Here, $p_{m}$ and $p_{DE}$ are the dust matter and dark energy pressures. In these equations, $Q$ is called the interaction term between dark matter and dark energy. This coupling between dark matter and dark energy may solve the coincidence problem in the $\Lambda CDM$ model \cite{bib48,bib49,bib50,bib51,bib52,bib53,bib54,bib55,bib56,bib57,bib58}.

On the DGP braneworld cosmology, $r_{c}$ stands for crossover length given by \cite{bib29,bib59}
\begin{eqnarray}\label{defin1}
r_{c}=\frac{M_{pl}^{2}}{2M_{5}^3}=\frac{G_{5}}{2G_{4}}.
\end{eqnarray}

Here $r_{c}$ has defined a length scale between the small and large distances. For the DGP scenario, $\epsilon$ shows the two separate branches of solutions, including $\epsilon= +1$ (or self-accelerating) branch and $\epsilon= -1$ (or normal) branch. The self-accelerating branch represents that the present cosmic acceleration happens without a dark energy component. However, the normal branch indicates that the acceleration needs to consider a dark energy ingredient.

For the spatially flat normal DGP braneworld, the Friedmann equation (2) reduces to
\begin{eqnarray}
H^{2}+\frac{H}{r_{c}}=\frac{1}{3M_{pl}^{2}}(\frac{\rho_{m}}{3}+\frac{\rho_{DE}}{3}).
\end{eqnarray}

The fractional energy densities will be
\begin{eqnarray}
\Omega_{m}=\frac{\rho_{m}}{3M_{pl}^{2}H^{2}} ,~\ \ \Omega_{DE}=\frac{\rho_{DE}}{3M_{pl}^{2}H^{2}},~\ \ \Omega_{r_{c}}=\frac{1}{4r_{c}^{2}H^{2}}.
\end{eqnarray}

Using the above definitions, we can rewrite the Friedmann equation as
\begin{eqnarray}
\Omega_{m}-2\sqrt{\Omega_{r_{c}}}+\Omega_{DE}=1.
\end{eqnarray}

The Tsallis agegraphic dark energy density is shown by
\begin{eqnarray}
\rho_{DE}=m T^{2\delta-4},
\end{eqnarray}
with
\begin{eqnarray}
T=\int_{0}^{a}dt=\int_{0}^{a}\frac{da}{Ha},
\end{eqnarray}
as IR cutoff. Also, from Eqs. (7) and (9), we have
\begin{eqnarray}
T=(\frac{3H^{2}\Omega_{DE}}{m})^{\frac{1}{2\delta-4}}.
\end{eqnarray}

The evolution of Tsallis agegraphic dark energy density will be
\begin{eqnarray}
\dot{\rho}_{DE}=\frac{3H^{2}\Omega_{DE}(2\delta-4)}{(\frac{3H^{2}\Omega_{DE}}{m})^{\frac{1}{2\delta-4}}},
\end{eqnarray}
in which, the $dot$ denotes the derivative with respect to cosmic time. From Eqs. (3), (6), (7), and (12) we can write the autonomous set of equations as
\begin{eqnarray}
&&\frac{d\Omega_{m}}{dN}=3b^{2}\Omega_{DE}+3\Omega_{m}(b^{2}-1)-2\Omega_{m}(\frac{\frac{dH}{dN}}{H}),\\
&&\frac{d\Omega_{DE}}{dN}=\frac{\Omega_{DE}(2\delta-4)}{H(\frac{3H^{2}\Omega_{DE}}{m})^{\frac{1}{2\delta-4}}}-2\Omega_{DE}(\frac{\frac{dH}{dN}}{H}),\\
&&\frac{d\Omega_{r_{c}}}{dN}=-2\Omega_{r_{c}}(\frac{\frac{dH}{dN}}{H}),
\end{eqnarray}
where $N=\ln(a)$, and
\begin{eqnarray}
&&\frac{dH}{dN}=\frac{H}{1+\sqrt{\Omega_{r_{c}}}}(\frac{3}{2}b^{2}\Omega_{DE}+\frac{3}{2}\Omega_{m}(b^{2}-1)\nonumber\\
&&+\frac{\Omega_{DE}(2\delta-4)}{2H(\frac{3H^{2}\Omega_{DE}}{m})^{\frac{1}{2\delta-4}}}).
\end{eqnarray}

Here, we have considered the simple choice of the interaction term between the dark sectors of the Universe
as \cite{bib60,bib61,bib62,bib63,bib64}
\begin{eqnarray}
Q=3b^{2}H(\rho_{DE}+\rho_{m}),
\end{eqnarray}
where $b^{2}$ is a coupling constant. \cite{bib65} obtained $b^{2}$ for ADE/NADE model for H(z), H(z)+CMB, and H(z)+CMB+BAO with 0.49/0.81, 0.04/0.09, and 0.16/0.39 values, respectively. \cite{bib66} assumed $b^{2}$ with 0, 0.02, 0.05, 0.10, and 0.15 values. Besides this, \cite*{bib67} considered $b^{2}$ with 0, 0.1, 0.2, and 0.3 values. In this work, in the first, we will assume $b^{2}$ with 0.00, 0.03, 0.05, and 0.07 with $\delta=2.5$ and $m=2.5$. Then we choose $b^{2}=0.03$ with supposed values of $\delta$ and $m$. After that, we calculate $b^{2}$ with the recent Hubble data.

Now, from Eq. (8), the equations (13), (14), and (15) can reduce to
\begin{eqnarray}
&&\frac{d\Omega_{DE}}{dN}=\frac{\Omega_{DE}(2\delta-4)}{H(\frac{3H^{2}\Omega_{DE}}{m})^{\frac{1}{2\delta-4}}}(1-\frac{\Omega_{DE}}{1+\sqrt{\Omega_{r_{c}}}}),
\nonumber\\ &&-\frac{3\Omega_{DE}}{1+\sqrt{\Omega_{r_{c}}}}(\Omega_{DE}b^{2}+(b^{2}-1)(1+2\sqrt{\Omega_{r_{c}}}-\Omega_{DE}))\\
&&\frac{d\Omega_{r_{c}}}{dN}=\frac{-3\Omega_{r_{c}}}{1+\sqrt{\Omega_{r_{c}}}}(\Omega_{DE}b^{2}+(b^{2}-1)(1\nonumber\\&&+2\sqrt{\Omega_{r_{c}}}-\Omega_{DE})+
\frac{\Omega_{DE}(2\delta-4)}{3H(\frac{3H^{2}\Omega_{DE}}{m})^{\frac{1}{2\delta-4}}}),
\end{eqnarray}
which the parameters $m$, $\delta$, and $b^{2}$ are the free parameters of the model.

\section{Cosmological Parameters}
\label{sec:2}

In this section, we will present the behavior of the model using the evolution of dark energy density ($\Omega_{DE}$), deceleration, equation of state, and total equation of state parameters.

First, we have plotted the evolution of $\Omega_{DE}$ versus redshift for different coupling $b^{2}$ (top plot), different values of $\delta$ (middle plot), and different values of $m$ (bottom plot) in figure 1. Also, we have taken $\Omega_{DE}(z=0)=0.680$, $H(z=0)=66.98$, and $\Omega_{r_{c}}=0.0003$ for the graphs.

\begin{figure*}
 \includegraphics[width=0.43\textwidth]{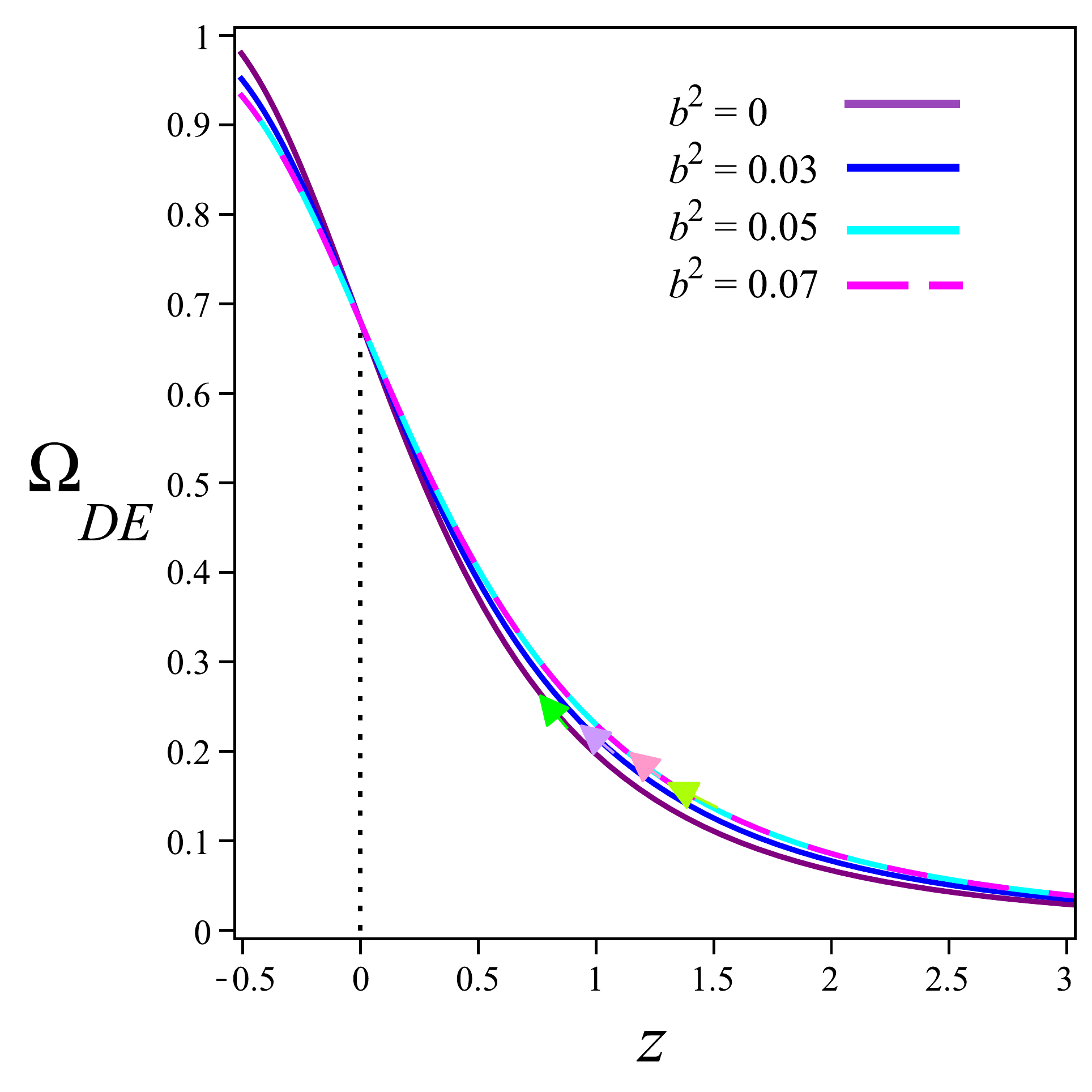}\\
 \includegraphics[width=0.43\textwidth]{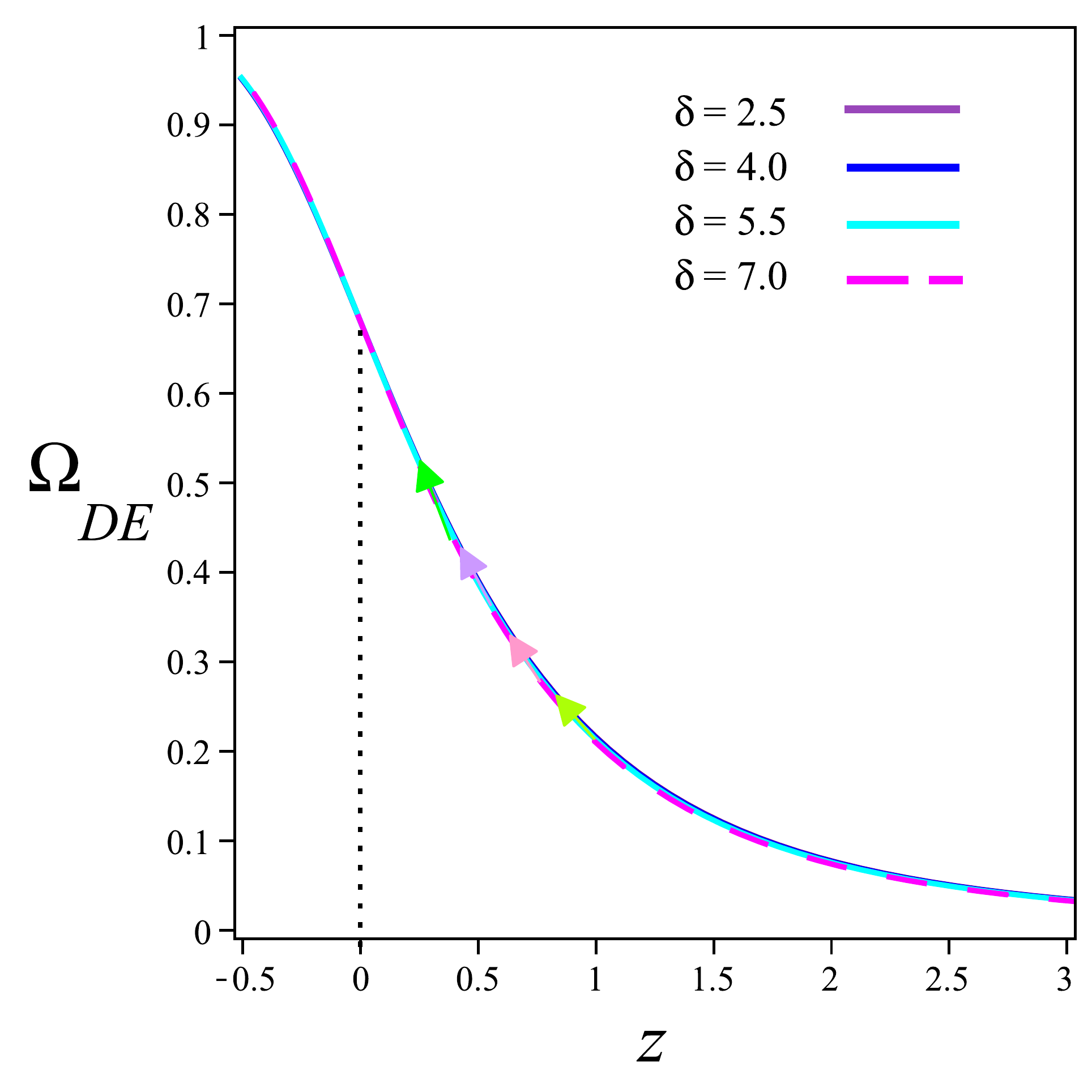}\\
 \includegraphics[width=0.43\textwidth]{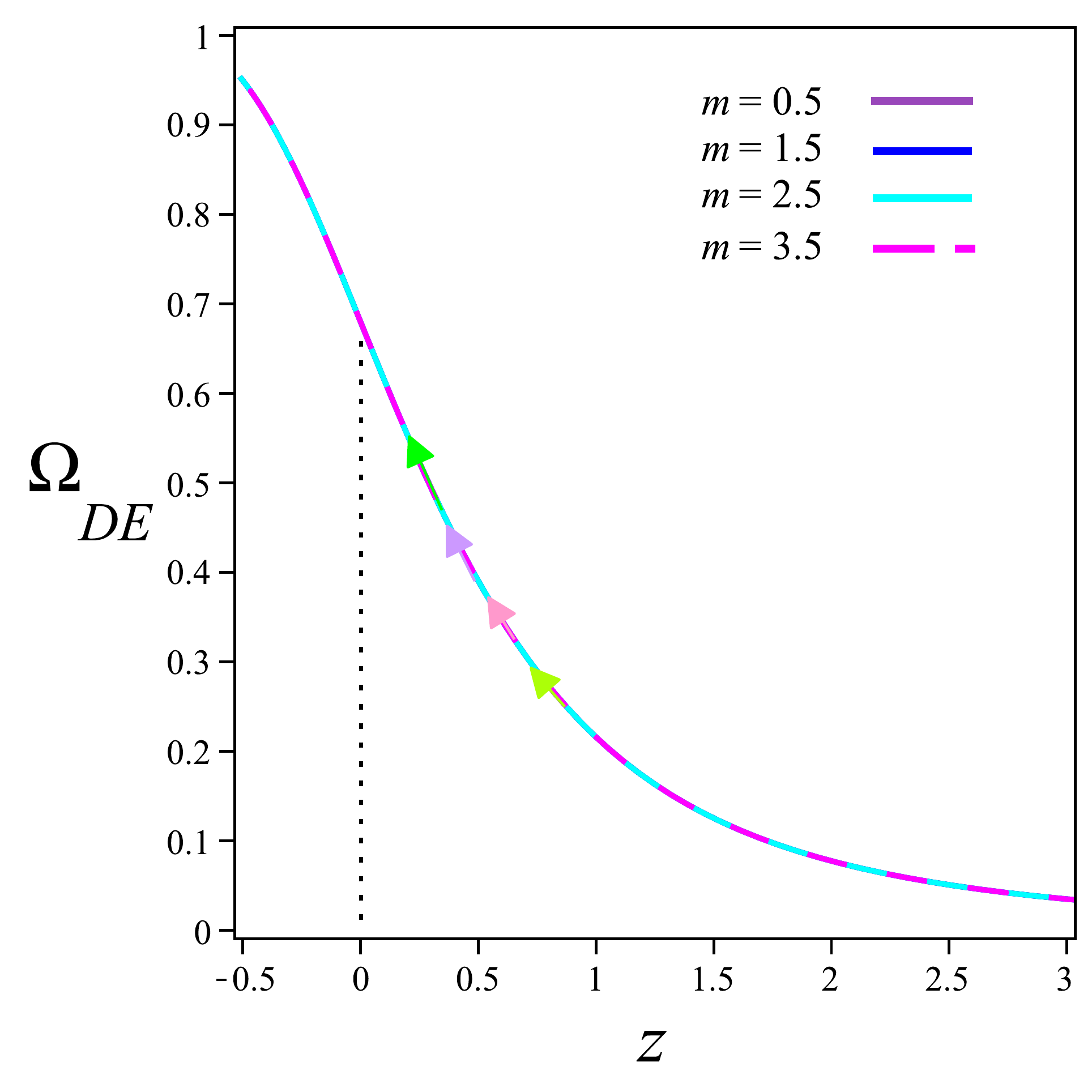}
 \caption{The evolution of $\Omega_{DE}$ versus redshift parameter z for the ITADE in the DGP braneworld cosmology. Here, we have taken $\Omega_{DE}(z=0)=0.680$, $H(z=0)=66.98$, and $\Omega_{r_{c}}=0.0003$ by
considering $\delta=2.5$, $m=2.5$ and different values of coupling $b^{2}$ (top plot), $b=0.03$, $m=2.5$ and different values of $\delta$ (middle plot), and $b=0.03$, $\delta=2.5$ and different values of $m$ (bottom plot)}
 \label{fig:1}       
\end{figure*}

This figure indicates that we have $\Omega_{DE}\longrightarrow 0$ at the early time ($z\longrightarrow\infty$) and $\Omega_{DE}\longrightarrow1$ at the future ($z\longrightarrow -0.5$).

The deceleration parameter is described by
\begin{eqnarray}\label{defin1}
q=-1-\frac{1}{H}\frac{dH}{dN}.\nonumber
\end{eqnarray}

Using Eq. (16), we find
\begin{eqnarray}
&&q=\frac{-1-\sqrt{\Omega_{r_{c}}}}{1+\sqrt{\Omega_{r_{c}}}}-\frac{\frac{3}{2}b^{2}\Omega_{DE}}{1+\sqrt{\Omega_{r_{c}}}}\nonumber\\&&-\frac{\frac{3}{2}(1+2\sqrt{\Omega_{r_{c}}}-\Omega_{DE})(b^{2}-1)}{1+\sqrt{\Omega_{r_{c}}}}-\frac{\frac{\Omega_{DE}(2\delta-4)}{2H(\frac{3H^{2}\Omega_{DE}}{m})^{\frac{1}{2\delta-4}}}}{1+\sqrt{\Omega_{r_{c}}}}.
\end{eqnarray}

If $q>0$, it shows the Universe is in the decelerating phase, but $q<0$ describes an accelerating phase of the Universe.
It is interesting to note that the observations suggest that the transition point from decelerating to accelerating time lies in the redshift range of $0.4<z<1$ (\cite{Gh14}).

In figure 2, we have plotted the evolution of the deceleration parameter for the ITADE model in the DGP braneworld scenario with respect to redshift for different coupling $b^{2}$, $\delta$, and $m$ considering $\Omega_{DE}(z=0)=0.680$, $H(z=0)=66.98$, and $\Omega_{r_{c}}=0.0003$.

\begin{figure*}
 \includegraphics[width=0.43\textwidth]{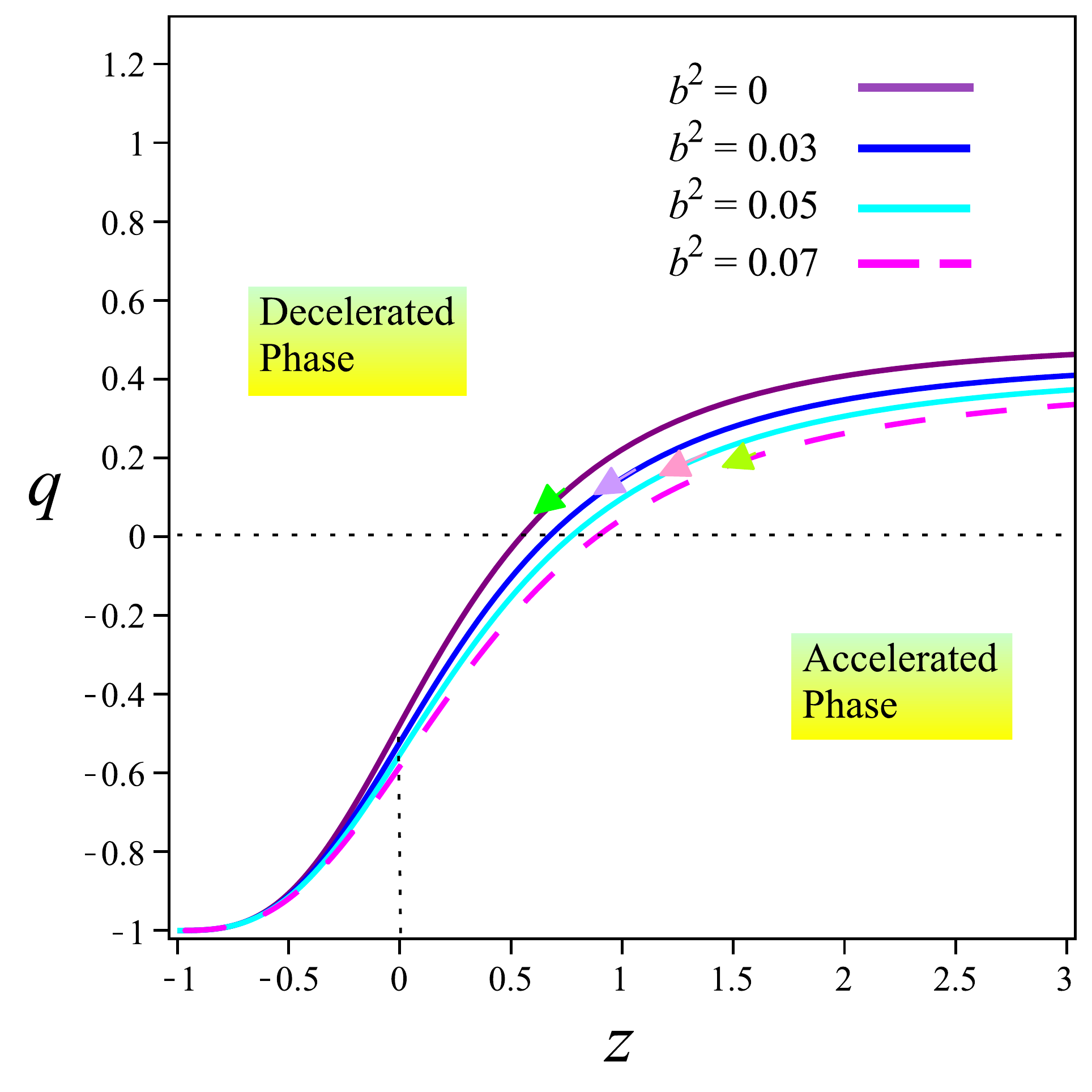}\\
 \includegraphics[width=0.43\textwidth]{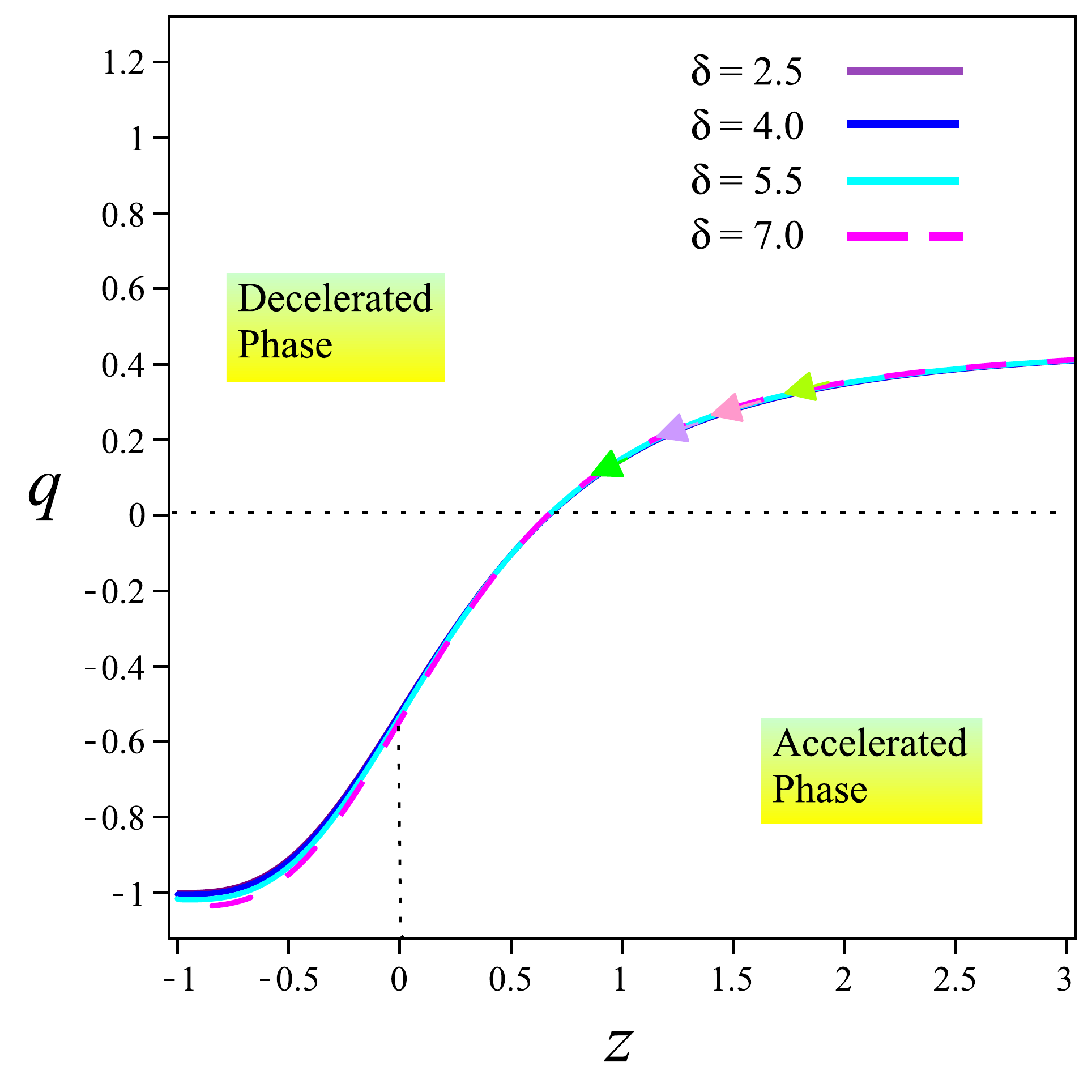}\\
\includegraphics[width=0.43\textwidth]{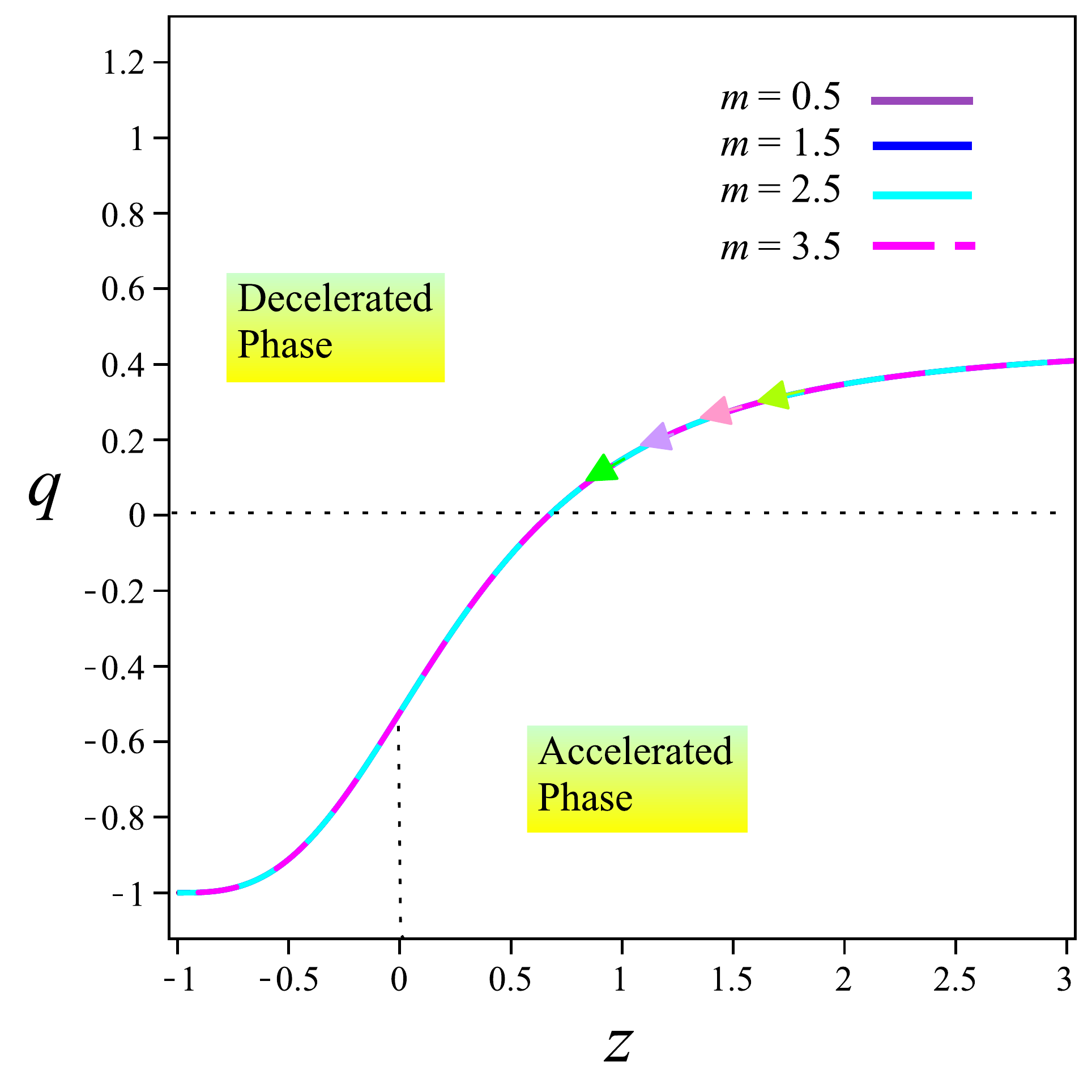}
 \caption{The evolution of q versus redshift parameter z for the ITADE in the DGP braneworld cosmology. Here, we have taken $\Omega_{DE}(z=0)=0.680$, $H(z=0)=66.98$, and $\Omega_{r_{c}}=0.0003$ by
  considering $\delta=2.5$, $m=2.5$ and different values of coupling $b^{2}$ (top plot), $b=0.03$, $m=2.5$ and different values of $\delta$ (middle plot), and $b=0.03$, $\delta=2.5$ and different values of $m$ (bottom plot)
}
 \label{fig:2}       
\end{figure*}

From Fig. 2, the Universe starts its acceleration at $0.547\leq z \leq0.886$, $0.666\leq z \leq0.671$, and $z= 0.671$ for different $b^{2}$, $\delta$, and $m$, respectively. We also see that the current value of $q$ is -0.477 (the solid purple line), -0.523 (the solid blue line), -0.553 (the solid cyan line), and -0.584 (the space dashed "Fuchsia" line) for various $b^{2}$. It is -0.523 (the solid purple line), -0.525 (the solid blue line), -0.534 (the solid cyan line), and -0.545 (the space dashed "Fuchsia" line) for various $\delta$, and -0.523 (for all lines) for various $m$.

The equation of state parameter can calculate using Eq. (4) and Eq. (12),
\begin{eqnarray}
&&w_{DE}=-1-b^{2}-(\frac{b^{2}+2b^{2}\sqrt{\Omega_{r_{c}}}-b^{2}\Omega_{DE}}{\Omega_{DE}})\nonumber\\&&
-\frac{2\delta-4}{3H(\frac{3H^{2}\Omega_{DE}}{m})^{\frac{1}{2\delta-4}}}.
\end{eqnarray}

It is important to note that $w_{DE}>-1$ shows the quintessence phase of dark energy candidate, while for $w_{DE}<-1$, it shows the phantom era. However, $w_{DE}=-1$ represents the cosmological constant barrier.

In Fig. 3, the dark energy behaves as the phantom at $z\longrightarrow\infty$ and $z\longrightarrow0$, while it approaches the $\Lambda$ state at $z\longrightarrow-1$ for different $b^{2}$, $\delta$, and $m$, except $b^{2}=0$. For $b^{2}=0$, the DE lies in the $w_{DE}=-1$ at $z\longrightarrow\infty$ and phantom phase at $z\longrightarrow0$ and $z\longrightarrow-1$ in the DGP model.

Also, from this figure we see that $w_{DE}(z=0)=-1.0000014,-1.046,-1.076$, and $-1.107$ for the top plot, $w_{DE}(z=0)=-1.046,-1.048,-1.057$, and $-1.068$ for the middle plot, while $w_{DE}(z=0)=-1.046$ (for the different values of $m$) in the bottom plot.

Apart from  $w_{DE}$, there is another equation of state parameter called $w_{tot}$ (the total equation of state parameter) described by
\begin{eqnarray}
&&w_{tot}=-1-\frac{2}{3}\frac{\frac{dH}{dN}}{H}=-\frac{1}{3}(1-2q),
\end{eqnarray}
or
\begin{eqnarray}
&&w_{tot}=\frac{p_{tot}}{\rho_{tot}}.
\end{eqnarray}

The total EoS parameter for the ITADE in the DGP braneworld model has been shown in Fig. 4. From this figure, we see that the Universe behaves as the quintessence at $z\longrightarrow\infty$ and $z\longrightarrow0$ for different $b^{2}$, $\delta$, and $m$. Then, it enters the phantom state for various $b^{2}$ and $m$, and $\delta= 2.5$, when $z\longrightarrow -1$. However, for $\delta= 4, 5.5$, and $7$ (in the middle plot), the Universe will approach -1 after entering the phantom region in the future.
Also, from figure 4, we can obtain the present values of $w_{tot}$, which are -0.651, -0.682, -0.702, and -0.723 in the top plane, -0.682, -0.684, -0.689, and -0.697 in the middle plane, and -0.682 (for all the lines) in the bottom plane.

Furthermore, comparing Figs. 3 and 4 show that although the $w_{DE}$ is always under the phantom divided line, the role of dark matter in the total energy density causes the total EoS parameter to lie in the quintessence era at the past and the present time.

\begin{figure*}
\includegraphics[width=0.43\textwidth]{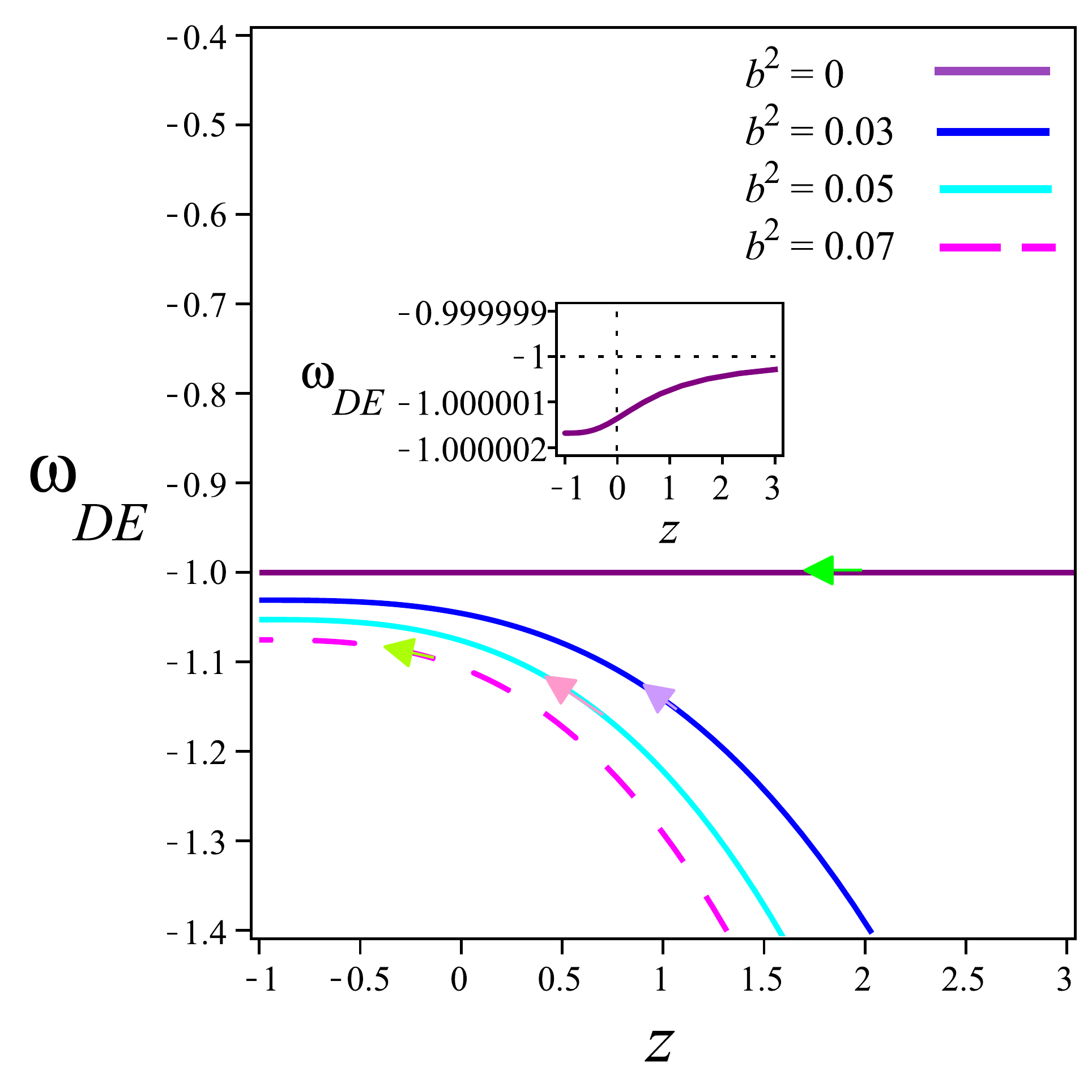}\\
 \includegraphics[width=0.43\textwidth]{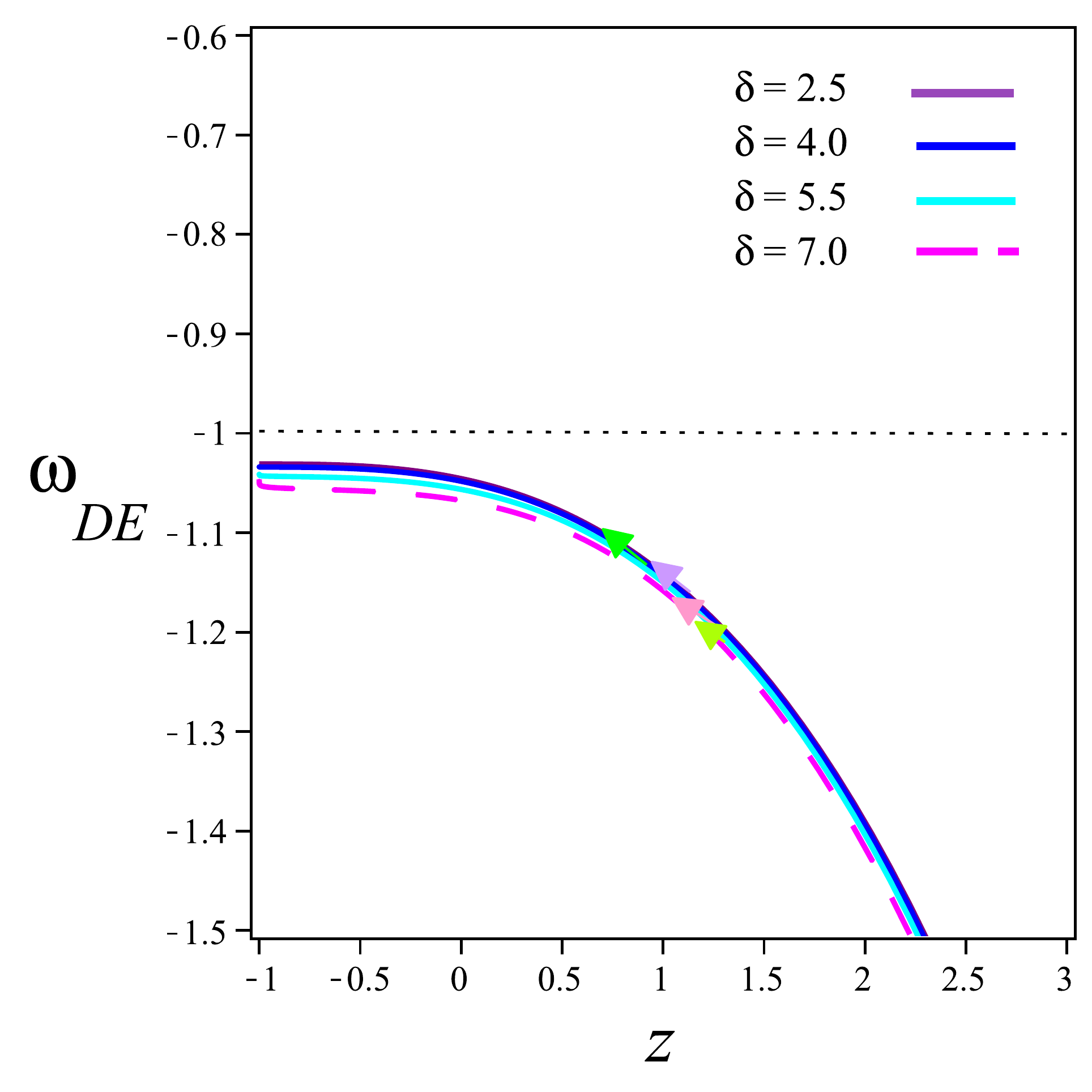}\\
 \includegraphics[width=0.43\textwidth]{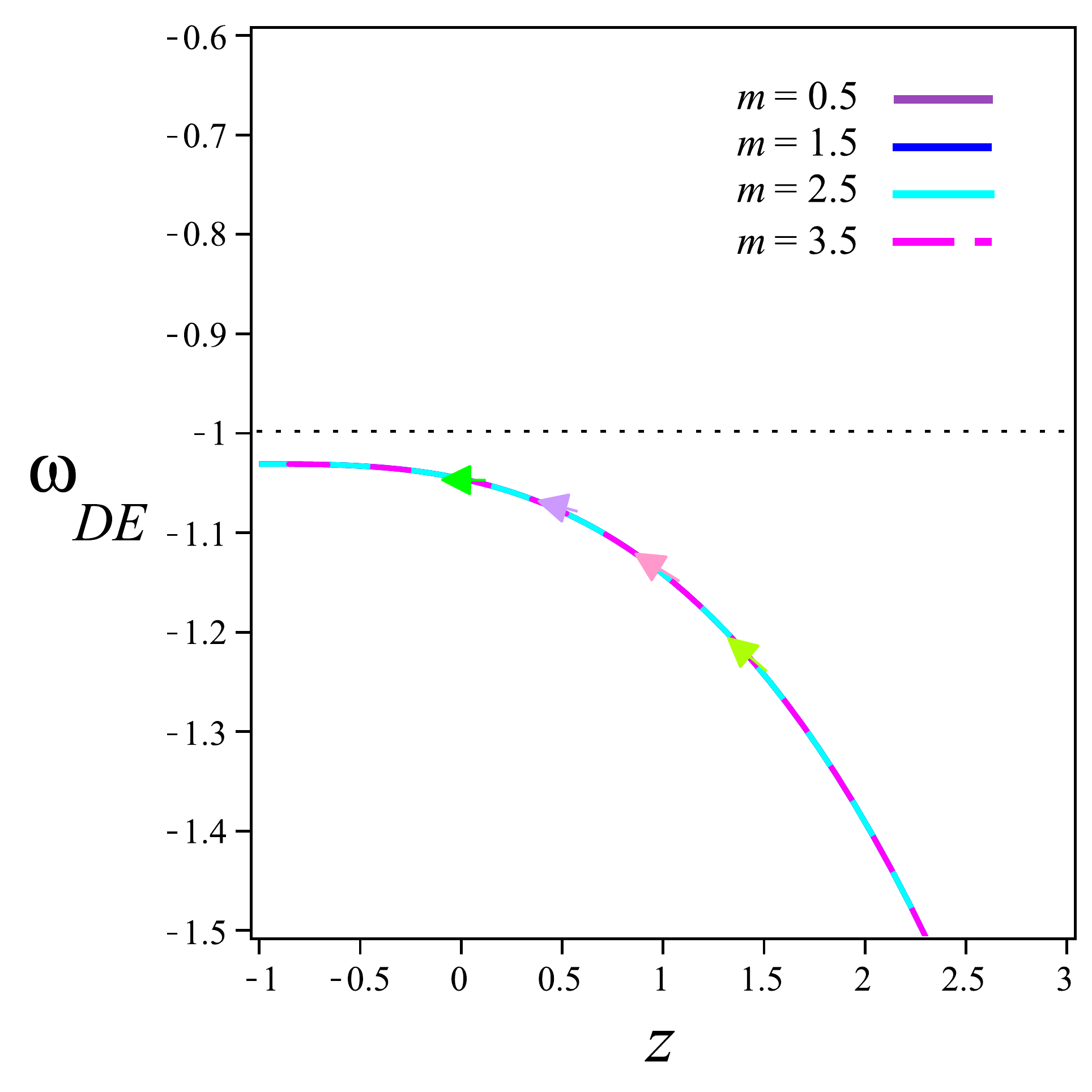}
 \caption{The evolution of $w_{DE}$ versus redshift parameter z for the ITADE in the DGP braneworld cosmology. Here, we have taken $\Omega_{DE}(z=0)=0.680$, $H(z=0)=66.98$, and $\Omega_{r_{c}}=0.0003$ by
  considering $\delta=2.5$, $m=2.5$ and different values of coupling $b^{2}$ (top plot), $b=0.03$, $m=2.5$ and different values of $\delta$ (middle plot), and $b=0.03$, $\delta=2.5$ and different values of $m$ (bottom plot)
  }
 \label{fig:3}       
\end{figure*}

\begin{figure*}
 \includegraphics[width=0.43\textwidth]{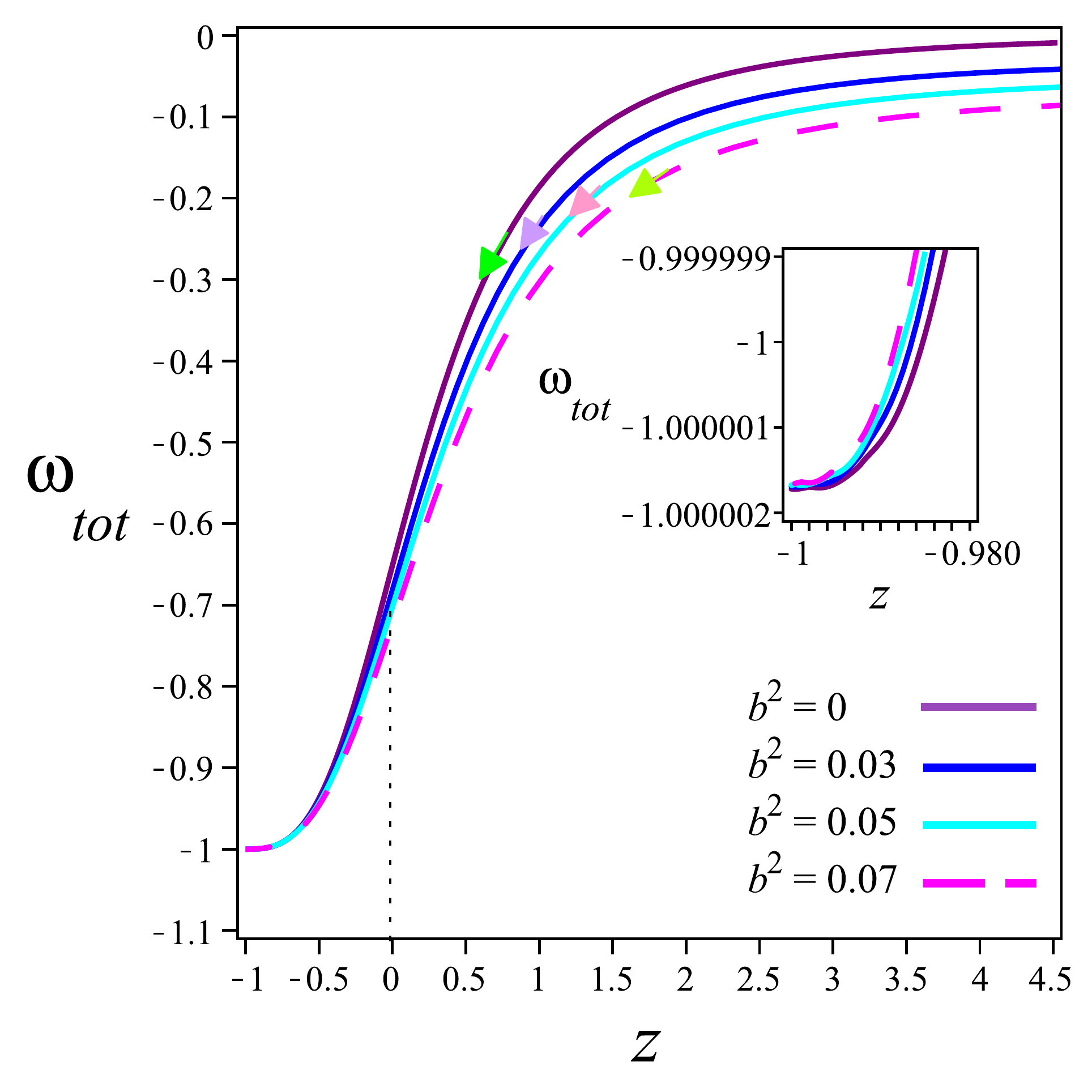}\\
\includegraphics[width=0.43\textwidth]{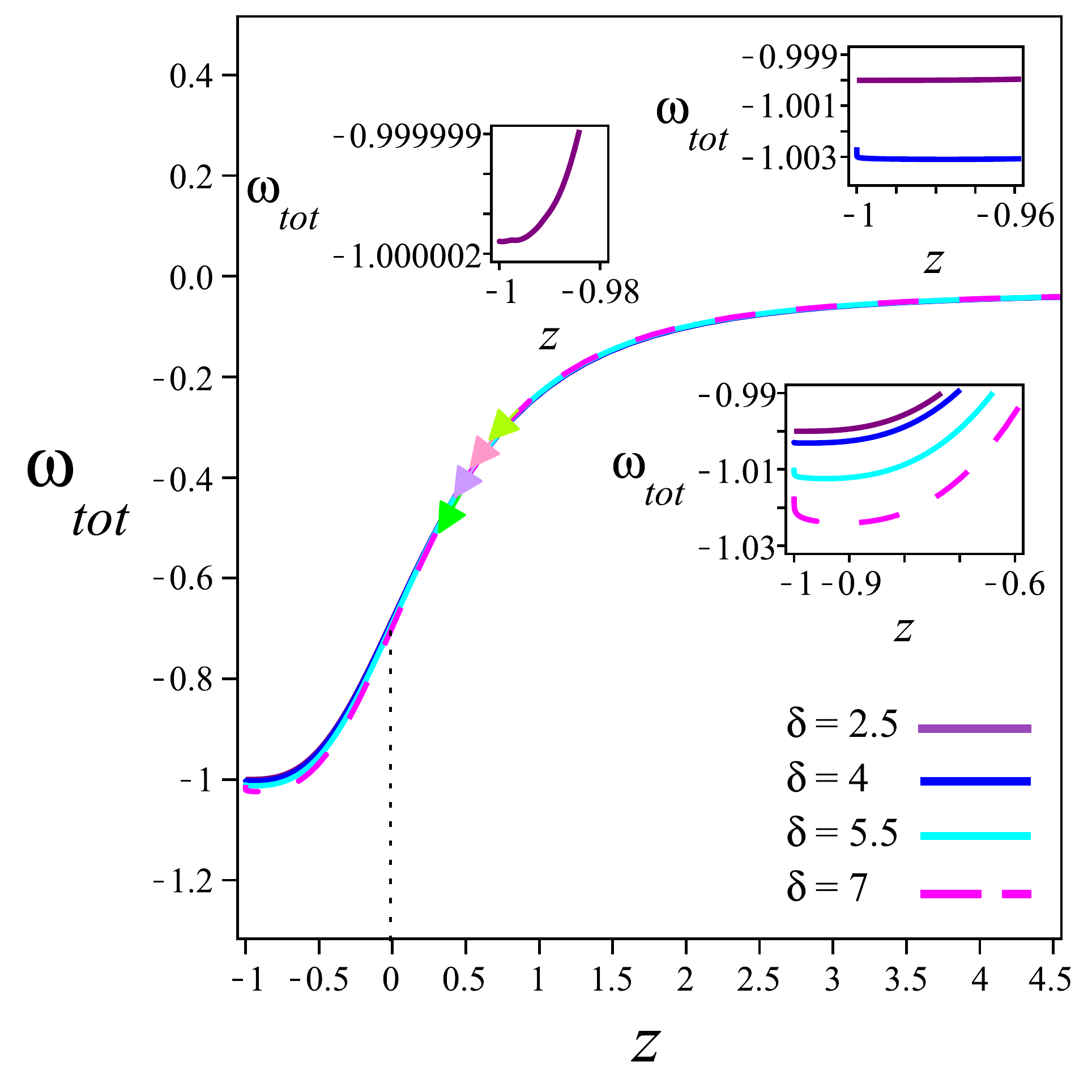}\\
\includegraphics[width=0.43\textwidth]{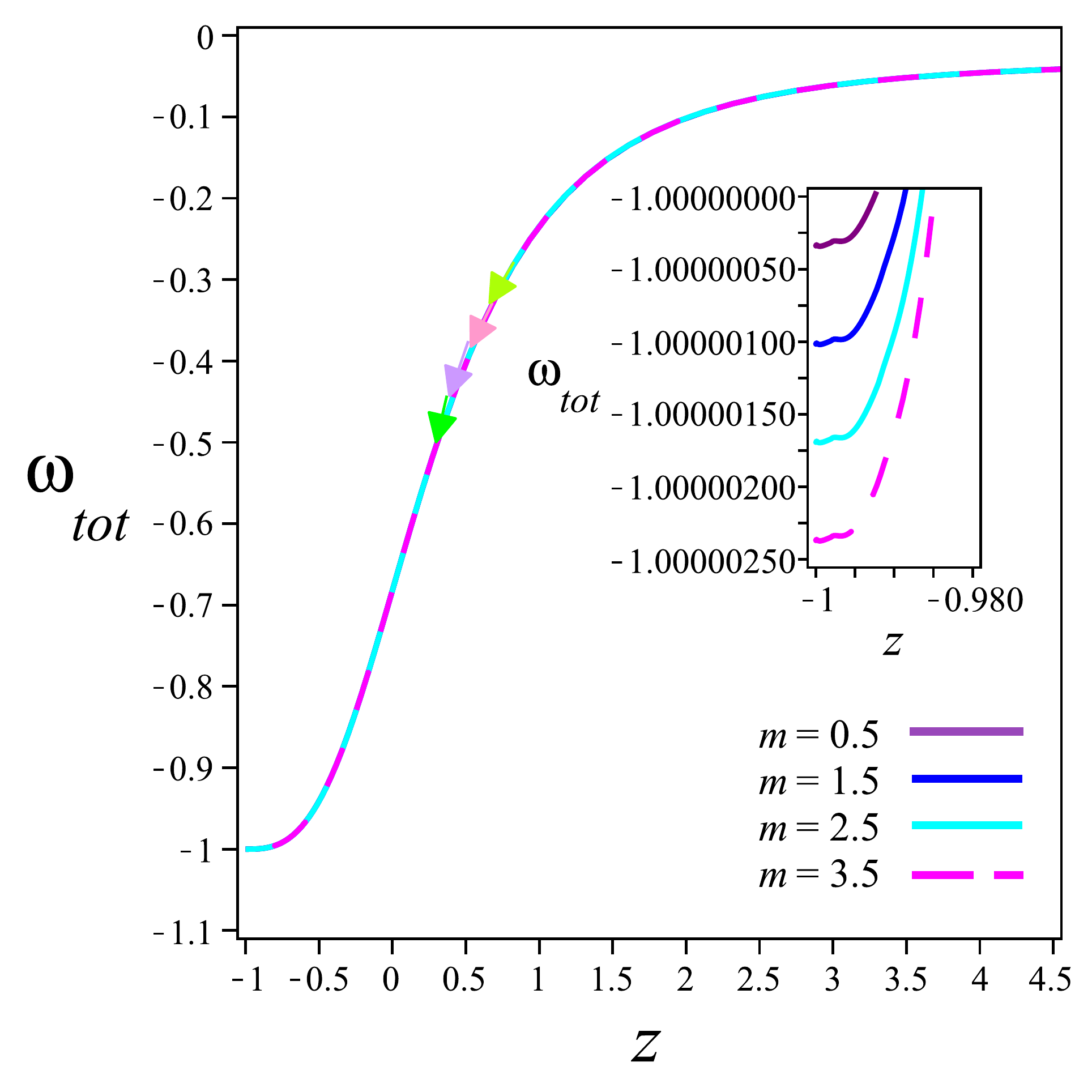}
 \caption{The evolution of $w_{tot}$ versus redshift parameter z for the ITADE in the DGP braneworld cosmology. Here, we have taken $\Omega_{DE}(z=0)=0.680$, $H(z=0)=66.98$, and $\Omega_{r_{c}}=0.0003$ by
  considering $\delta=2.5$, $m=2.5$ and different values of coupling $b^{2}$ (top plot), $b=0.03$, $m=2.5$ and different values of $\delta$ (middle plot), and $b=0.03$, $\delta=2.5$ and different values of $m$ (bottom plot)
}
 \label{fig:4}       
\end{figure*}

\section{Statefinder Diagnostic}
\label{sec:3}

In this section, we study the ITADE model in the DGP braneworld scenario through the statefinder pairs, especially $\{r-s\}$ (\cite*{Sahn3}). This plane can discriminate between different dark energy models. The statefinder parameters r and s can be described by:
\begin{eqnarray}
r=2q^{2}+q-\frac{dq}{dN},
\end{eqnarray}

\begin{eqnarray}
s=\frac{r-1}{3(q-\frac{1}{2})}.
\end{eqnarray}

Fig. 5 shows the evolutionary trajectory of $\{r-s\}$ for the model considering different $b^{2}$, $\delta$, and $m$. In figure 5, the fixed point values of statefinder parameters $\{r, s\}= \{1,0\}$ (the solid "HotPink" circles) and $\{r, s\}= \{1,1\}$ (the solid "MediumOrchid" circles) represent the $\Lambda CDM$ and SCDM models, respectively.
From this figure, the region of the chaplygin gas ($s<0, r>1$) and the quintessence ($s>0, r<1$) can be observed as well. Checking tracks of the panels indicate that the model behaves as the quintessence at the past and present time, and then it will tend to the $\Lambda CDM$ in the future era.
The "Orange" circle in the top and middle panels relates to the current values of the statefinder parameters with $\{r_{0}, s_{0}\}\longrightarrow \{-0.044, 0.356\}, \{0.033, 0.315\}, \{0.085, 0.290\}$, and\\ $\{0.139, 0.265\}$ for $b^{2}$, and $\{r_{0}, s_{0}\}\longrightarrow \{0.033, 0.315\},\\ \{0.047, 0.310\}$, $\{0.092, 0.293\}$, and $\{0.153, 0.270\}$ for $\delta$. Also, in this figure, the solid "Gold" circle in the bottom plane indicates the $\{r_{0}, s_{0}\}\longrightarrow \{0.033, 0.315\}$ (for all values of $m$).

\begin{figure*}
\includegraphics[width=0.43\textwidth]{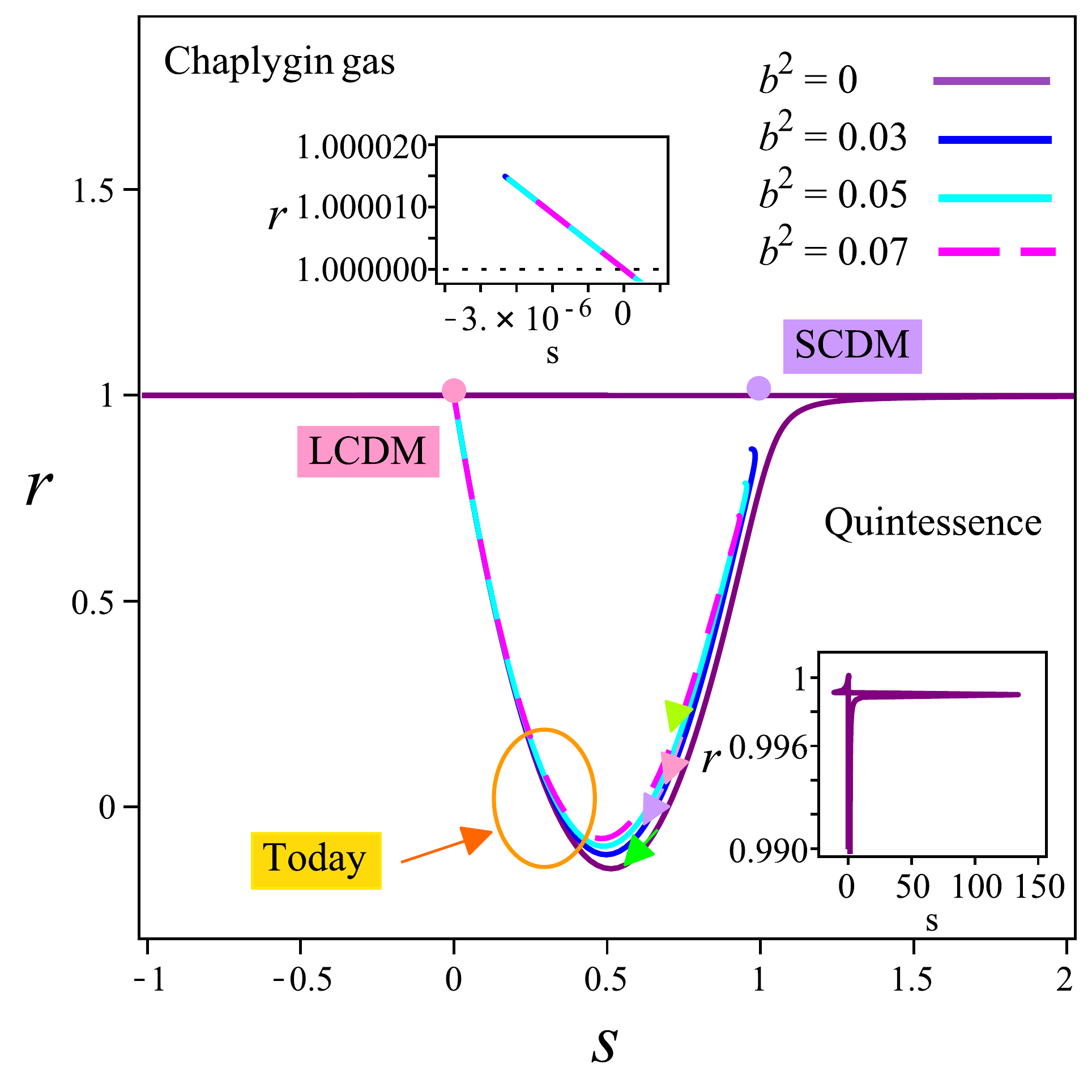}\\
 \includegraphics[width=0.43\textwidth]{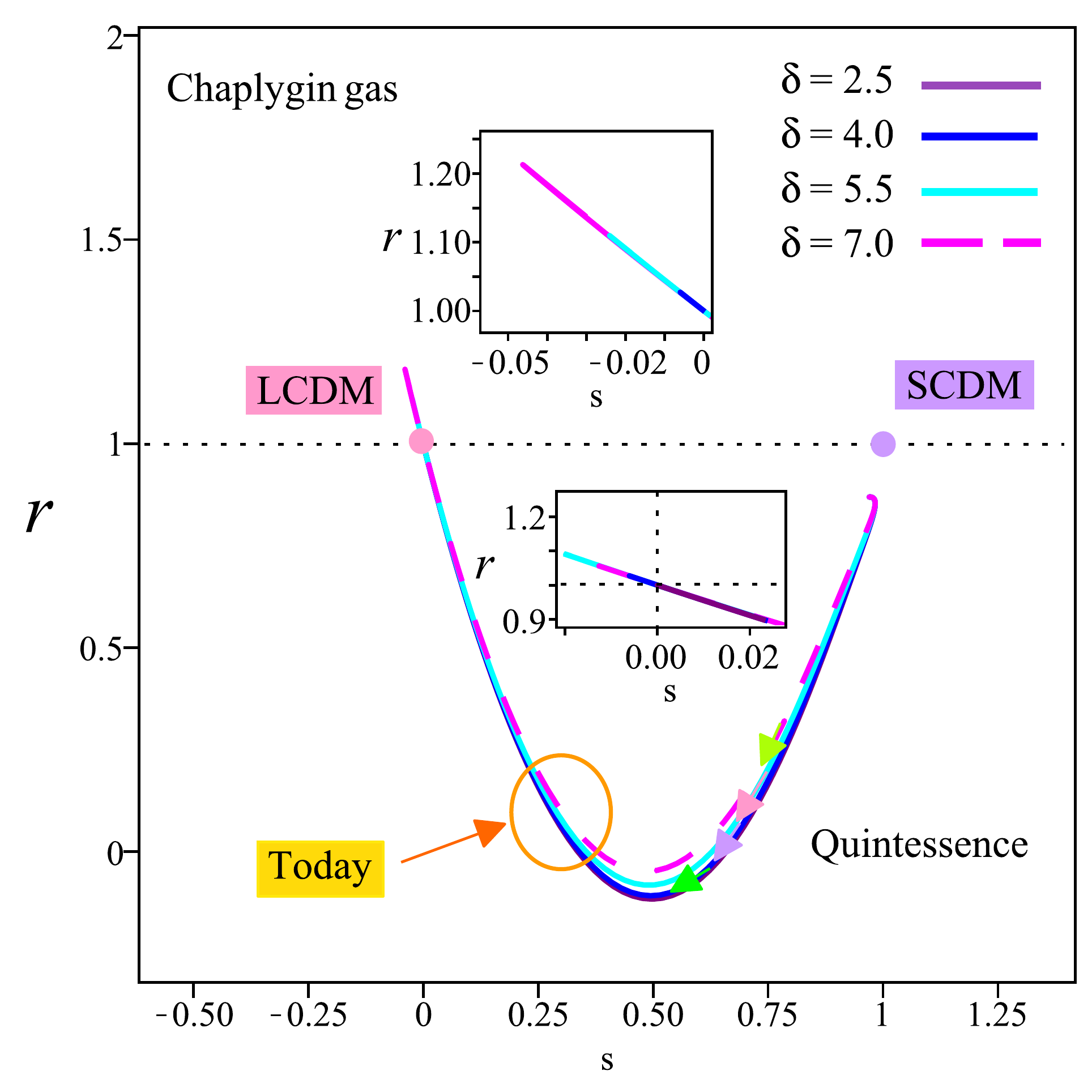}\\
\includegraphics[width=0.43\textwidth]{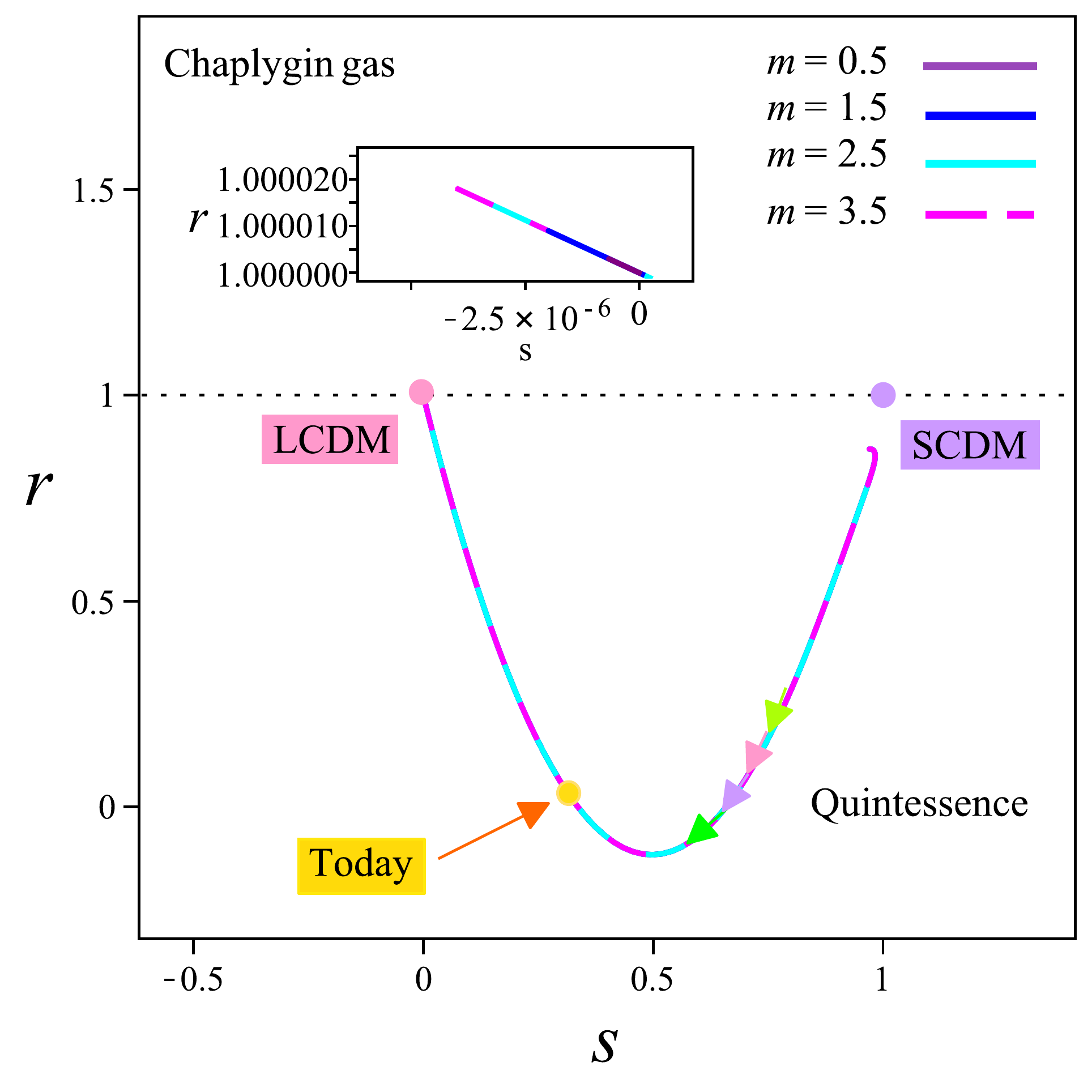}
 \caption{The evolution of r versus s for the ITADE in the DGP braneworld cosmology. Here, we have taken $\Omega_{DE}(z=0)=0.680$, $H(z=0)=66.98$, and $\Omega_{r_{c}}=0.0003$ by
  considering $\delta=2.5$, $m=2.5$ and different values of coupling $b^{2}$ (top plot), $b=0.03$, $m=2.5$ and different values of $\delta$ (middle plot), and $b=0.03$, $\delta=2.5$ and different values of $m$ (bottom plot)
}
 \label{fig:5}       
\end{figure*}

\section{The $\omega^{'}_{DE}-\omega_{DE}$ Analysis}
\label{sec:4}

\cite*{Caldw5} proposed the $\omega^{'}_{DE}-\omega_{DE}$ plane to explain the dynamical properties of DE models. This plane describes the quintessence phase of DE divided into two different classes known as the thawing region ($\omega^{'}_{DE}>0, \omega_{DE}<0$) and freezing region ($\omega^{'}_{DE}<0, \omega_{DE}<0$). The thawing region explains the Universe does not support a permanent acceleration, while the freezing part indicates the field provides a sustained acceleration.

Here, we calculate $\omega^{'}_{DE}$ from Eq. (21) as
\begin{eqnarray}
&&\omega^{'}_{DE}=-\frac{b^{2}\frac{d\Omega_{r_{c}}}{dN}}{\Omega_{DE}\sqrt{\Omega_{r_{c}}}}
+\frac{b^{2}\frac{d\Omega_{DE}}{dN}}{\Omega_{DE}} \nonumber \\&&
 +\frac{b^{2}\frac{d\Omega_{DE}}{dN}(1+2\sqrt{\Omega_{r_{c}}}-\Omega_{DE})}{\Omega_{DE}^{2}}
 \nonumber \\&&+\frac{2\delta-4}{3H^{2}(\frac{3H^{2}\Omega_{DE}}{m})^{\frac{1}{2\delta-4}}}\frac{dH}{dN}
+\frac{2}{3H^{2}(\frac{3H^{2}\Omega_{DE}}{m})^{\frac{1}{2\delta-4}}}\frac{dH}{dN}
 \nonumber \\&&+\frac{\frac{d\Omega_{DE}}{dN}}{3H\Omega_{DE}(\frac{3H^{2}\Omega_{DE}}{m})^{\frac{1}{2\delta-4}}}.
\end{eqnarray}

\begin{figure*}
\includegraphics[width=0.43\textwidth]{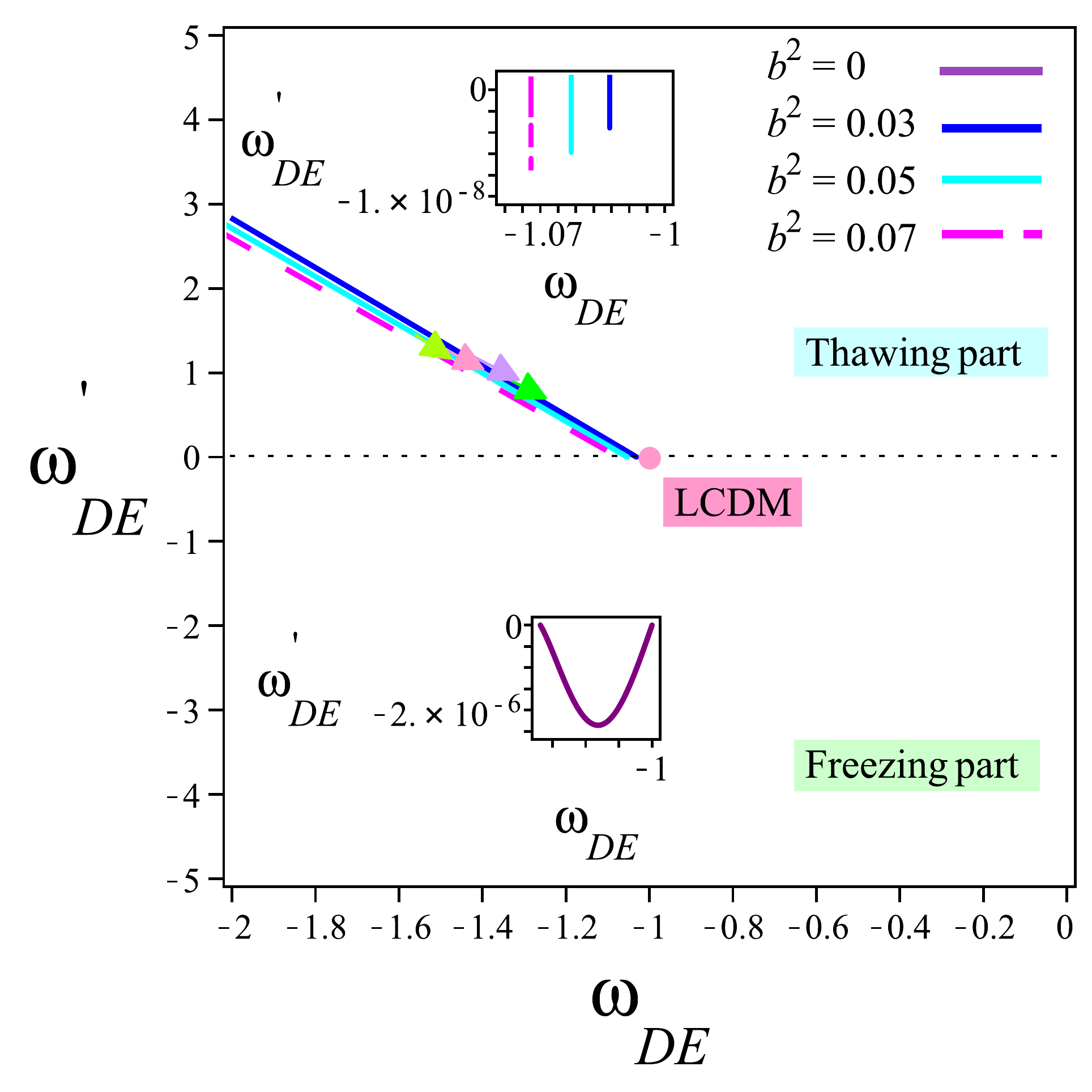}\\
 \includegraphics[width=0.43\textwidth]{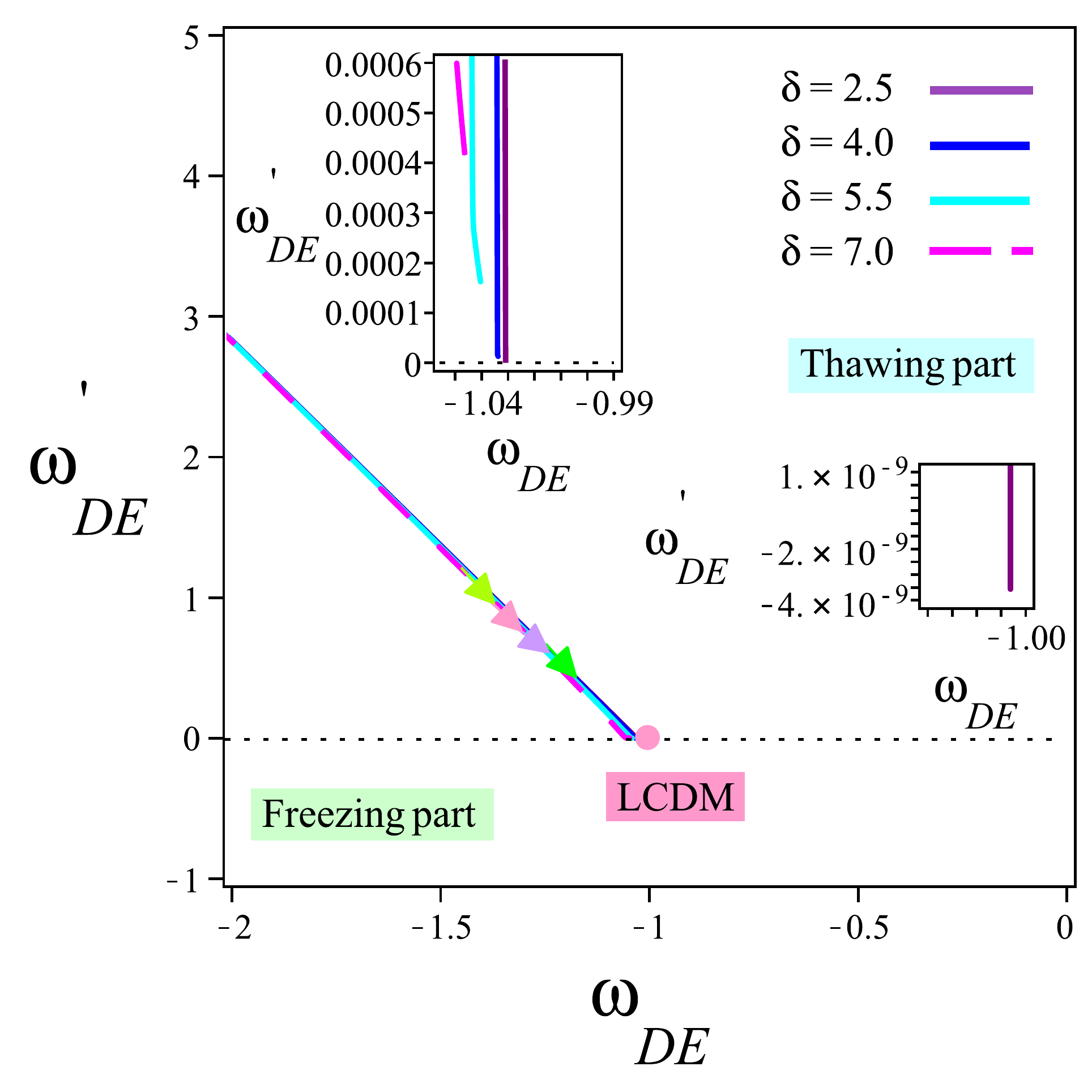}\\
 \includegraphics[width=0.43\textwidth]{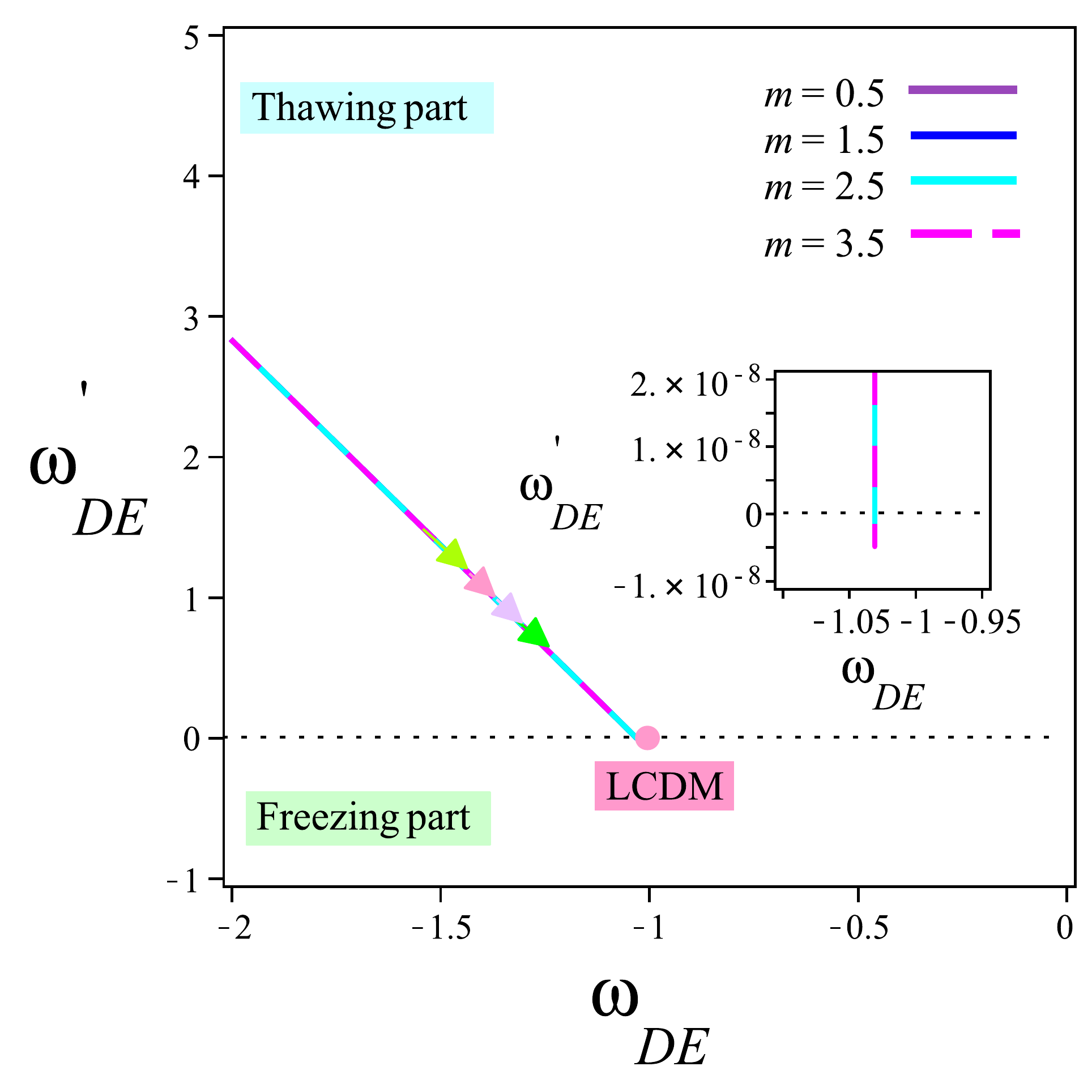}
 \caption{The $\omega^{'}_{DE}-\omega_{DE}$ diagram for the ITADE in the DGP braneworld cosmology.  Here, we have taken $\Omega_{DE}(z=0)=0.680$, $H(z=0)=66.98$, and $\Omega_{r_{c}}=0.0003$ by
  considering $\delta=2.5$, $m=2.5$ and different values of coupling $b^{2}$ (top plot), $b=0.03$, $m=2.5$ and different values of $\delta$ (middle plot), and $b=0.03$, $\delta=2.5$ and different values of $m$ (bottom plot)
 }
 \label{fig:6}       
\end{figure*}

In Fig. 6, we have plotted the $\omega^{'}_{DE}-\omega_{DE}$ plane for the ITADE model in the DGP scenario for different $b^{2}$, $\delta$, and $m$. In this figure, the solid "HotPink" circle corresponds to the $\Lambda CDM$ fixed point with $\omega^{'}_{DE}=0$ and $\omega_{DE}=-1$. From Fig. 6 (top plane), we observe that the model for the non-interaction term ($b^{2}=0$) experienced the $\Lambda CDM$ state in the far past, the freezing phase in the past and present, while it will go toward the thawing part in the future. For $b^{2}= 0.03, 0.05$, and $0.07$, we have the thawing part for the past and present and the freezing region for the future.
Then, from the middle plane of Fig. 6, the thawing feature can be recognized in the past, present, and future for different values of $\delta$. However, $\delta= 2.5$ will experience freezing behavior in the future. Finally, the bottom plane of Fig. 6 indicates the thawing part of the model for the past and current time and freezing state in the future.\\\\\\\\\\\\

\section{Stability}
\label{sec:5}
To explore the viability of the ITADE model in the DGP braneworld cosmology, we investigate the stability of the model against small perturbation using the sign of square of the sound speed $v_{s}^{2}$ parameter. The adiabatic sound speed is given by
\begin{eqnarray}\label{defin1}
v_{s}^{2}=\frac{dP_{DE}}{d\rho_{DE}}=H\frac{\rho_{DE}}{\dot{\rho}_{DE}} \omega_{DE}^{'}+\omega_{DE}.\nonumber
\end{eqnarray}

The signs of $v_{s}^{2}$ denote a stable ($v_{s}^{2}>0$) and an unstable ($v_{s}^{2}<0$) Universe against perturbation.

\begin{figure*}
\includegraphics[width=0.43\textwidth]{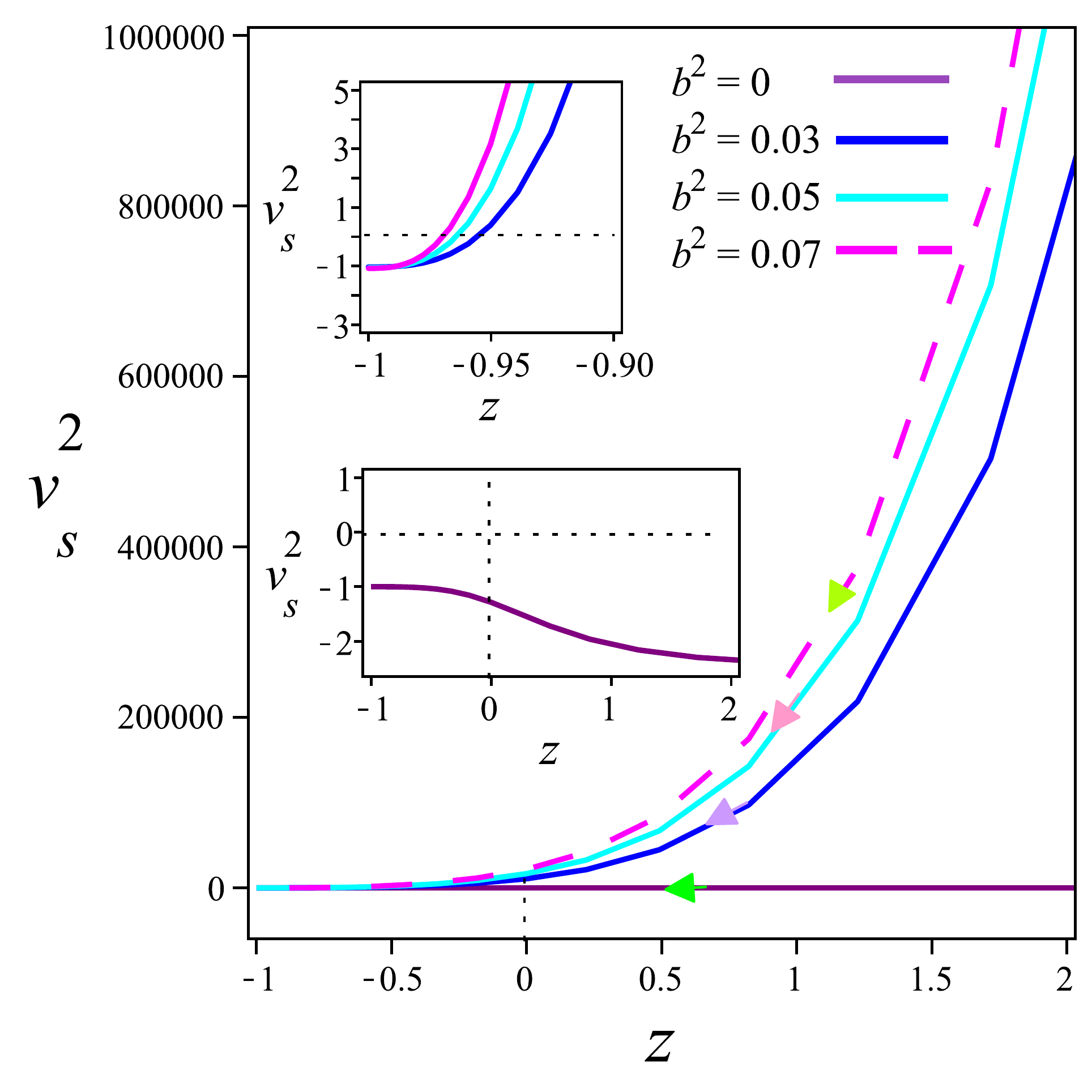}\\
\includegraphics[width=0.43\textwidth]{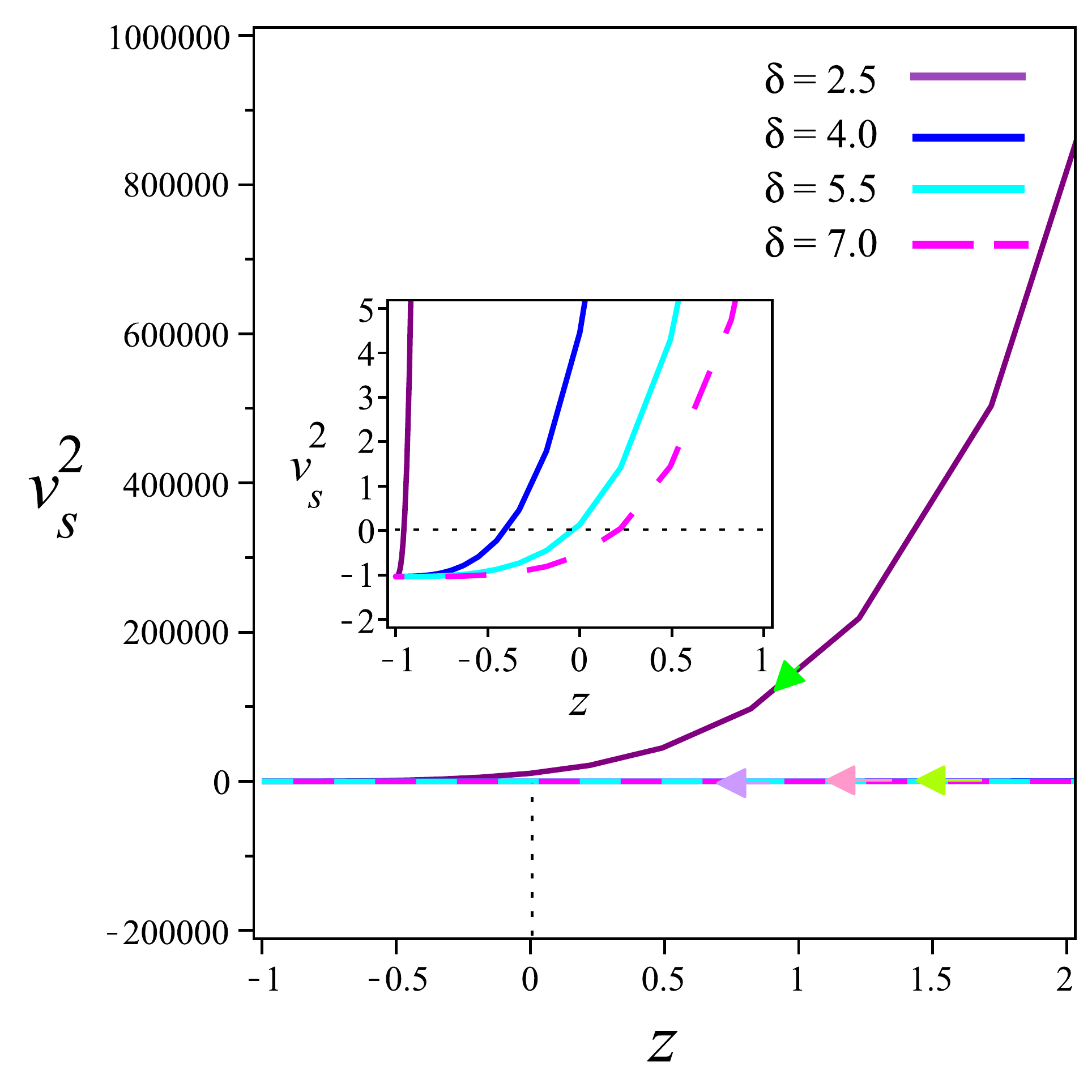}\\
\includegraphics[width=0.43\textwidth]{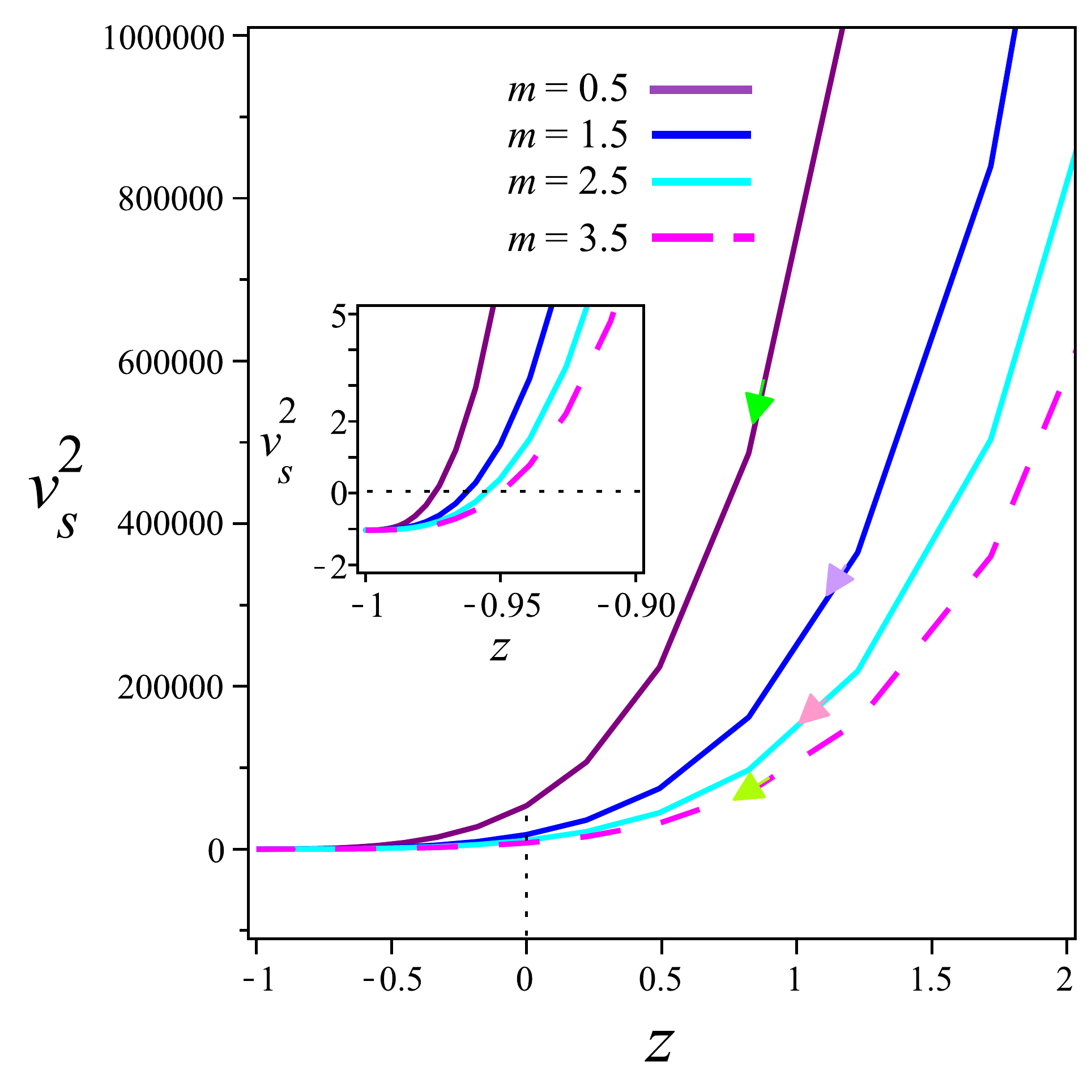}
\small
 \caption{The evolution of $v_{s}^{2}$ versus redshift parameter z for the ITADE in the DGP braneworld cosmology. Here, we have taken $\Omega_{DE}(z=0)=0.680$, $H(z=0)=66.98$, and $\Omega_{r_{c}}=0.0003$ by
  considering $\delta=2.5$, $m=2.5$ and different values of coupling $b^{2}$ (top plot), $b=0.03$, $m=2.5$ and different values of $\delta$ (middle plot), and $b=0.03$, $\delta=2.5$ and different values of $m$ (bottom plot)
 }
 \label{fig:7}       
\end{figure*}

In figure 7, we have plotted the evolution of the adiabatic sound speed parameter for the ITADE in the DGP scenario against redshift for different $b^{2}$, $\delta$, and $m$ considering $\Omega_{DE}(z=0)=0.680$, $H(z=0)=66.98$, and $\Omega_{r_{c}}=0.0003$.

Clearly, we see that $v_{s}^{2}>0$ at $z\longrightarrow\infty$ for the top, middle, and bottom graphs except for the purple line in the top graph corresponded to the non-interaction term ($b^{2}=0$). For $z\longrightarrow 0$ these graphs show stability for different $b^{2}$, $\delta$, and $m$. Although, $b^{2}=0$ and $\delta=7$ exhibit instability situations for the Universe. Finally, From Fig. 7, it is obvious that the Universe will be unstable for considered values of $b^{2}$, $\delta$, and $m$ for $z\longrightarrow -1$.

\section{Test the model with Hubble data set}
\label{sec:6}

Now, we are planning to fit the parameters of the model in the system of Eqs. (18) and (19) by using Hubble observational data.

The $\chi^{2}_{Hub}$ for the Hubble parameter is described by
\begin{eqnarray}\label{chisquare}
\chi_{min}^{2}=\sum_{i=1}^{30}\frac{[H_{i}^{th}-H_{i}^{obs}]^{2}}{\sigma_{i}^{2}}.
\end{eqnarray}

The best-fitted parameters, $\Omega_{DE0}$, $\Omega_{r_{c}0}$, $\delta$, $m$, and $b$, are shown in Table 1. In addition, we have provided the best fit values of the $\Lambda CDM$ model (in Table 1) to compare with our present work. For this reason, we prefer to use the Akaike Information Criterion (AIC) (\cite*{bib70}) and Bayesian Information Criterion (BIC) (\cite{bib71}) methods.
The AIC and BIC describe as
\begin{eqnarray}
AIC=\chi_{min}^{2}+2K,\nonumber
\end{eqnarray}
and
\begin{eqnarray}
BIC=\chi_{min}^{2}+K \ln N.\nonumber
\end{eqnarray}

Here $K$ is the number of free parameters, and $N$ is the number of data points. To compare model 1 with model 2, we calculate
\begin{eqnarray}
\Delta AIC= AIC_{1}-AIC_{2},\nonumber
\end{eqnarray}
and
\begin{eqnarray}
\Delta BIC= AIC_{1}-AIC_{2}.\nonumber
\end{eqnarray}

For $0\leq\Delta AIC<2$, $4<\Delta AIC\leq7$, and $8\leq\Delta AIC>10$, we have strong evidence, little evidence, and no evidence in favor of the model 1, while $0\leq\Delta BIC<2$, $2\leq\Delta BIC\leq6$, $6\leq\Delta BIC<10$, and $BIC>10$ shows not enough evidence, evidence, strong evidence, and very strong evidence against model 1.

In Table 1, we have summarized the results for the present model and the $\Lambda CDM$ model parameters. The AIC result shows that the ITADE in the DGP braneworld is less supported by the $\Lambda CDM$ model and observational data. The result of the BIC shows the ITADE in the DGP braneworld is relatively very strong evidence against the cosmological model. However, it solves the $\Lambda CDM$ difficulties.

\begin{table}[t]
\caption{The best-fit values for the model parameters}
\begin{tabular}{@{}llll}
\hline
Parameter & ITADE & $\Lambda CDM$\\
\hline
$H_{0}$~~ &$fixed$ $(66.98)$ &$68.13$\\
\hline
$\Omega_{DE0}$~~ &$0.599$  &$0.680$\\
\hline
$\Omega_{r_{c}0}$~~ &$0.0018$  &--\\
\hline
$m$ &$3.0$   &--\\
\hline
$\delta$ &$12.7$   &--\\
\hline
$b$ &$0.28$  &--\\
\hline
$\chi^{2}_{min}$ &$\simeq 14.600$  &14.516\\
\hline
$\chi^{2}_{dof}$ &$0.584$   &0.518\\
\hline
AIC  &24.6  &18.516\\
\hline
$\Delta AIC$ &6.084  &0\\
\hline
BIC & $\simeq 31.606$  &$\simeq21.318$ \\
\hline
$\Delta BIC$ &10.288  & 0\\
\hline
\end{tabular}
\end{table}

In Fig. 8, we have presented the constraints on the parameters $\Omega_{DE}$, $\delta$, and $b$ at the 68.3$\%$ and 95.4$\%$ confidence levels. Then, in Fig. 9, we have plotted 1-dim and 2-dim likelihood distribution for $\Omega_{r_{c}}$-$\Omega_{DE}$, $\delta$, and $m$ parameters.

\begin{figure*}
\includegraphics[width=0.43\textwidth]{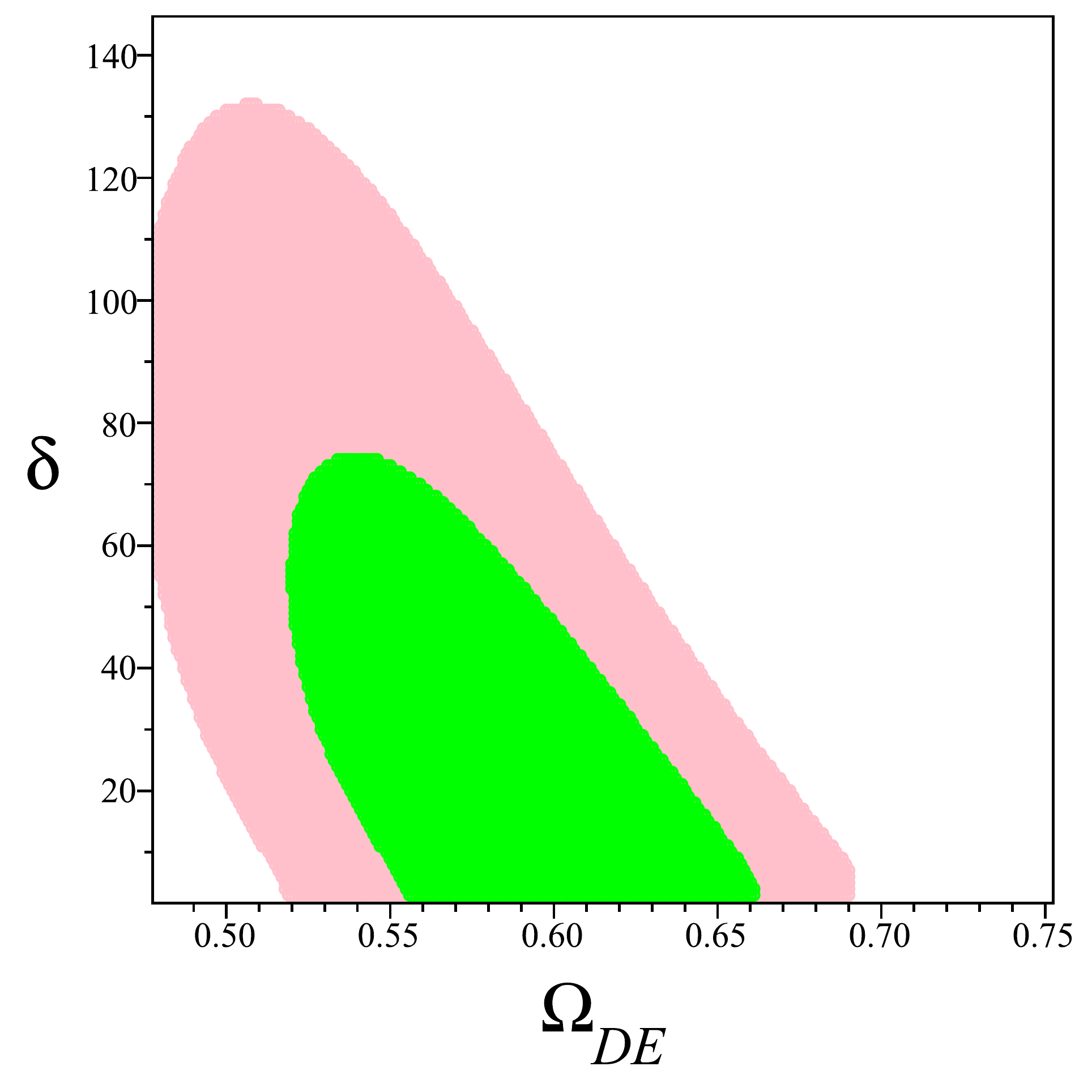} \\
\includegraphics[width=0.43\textwidth]{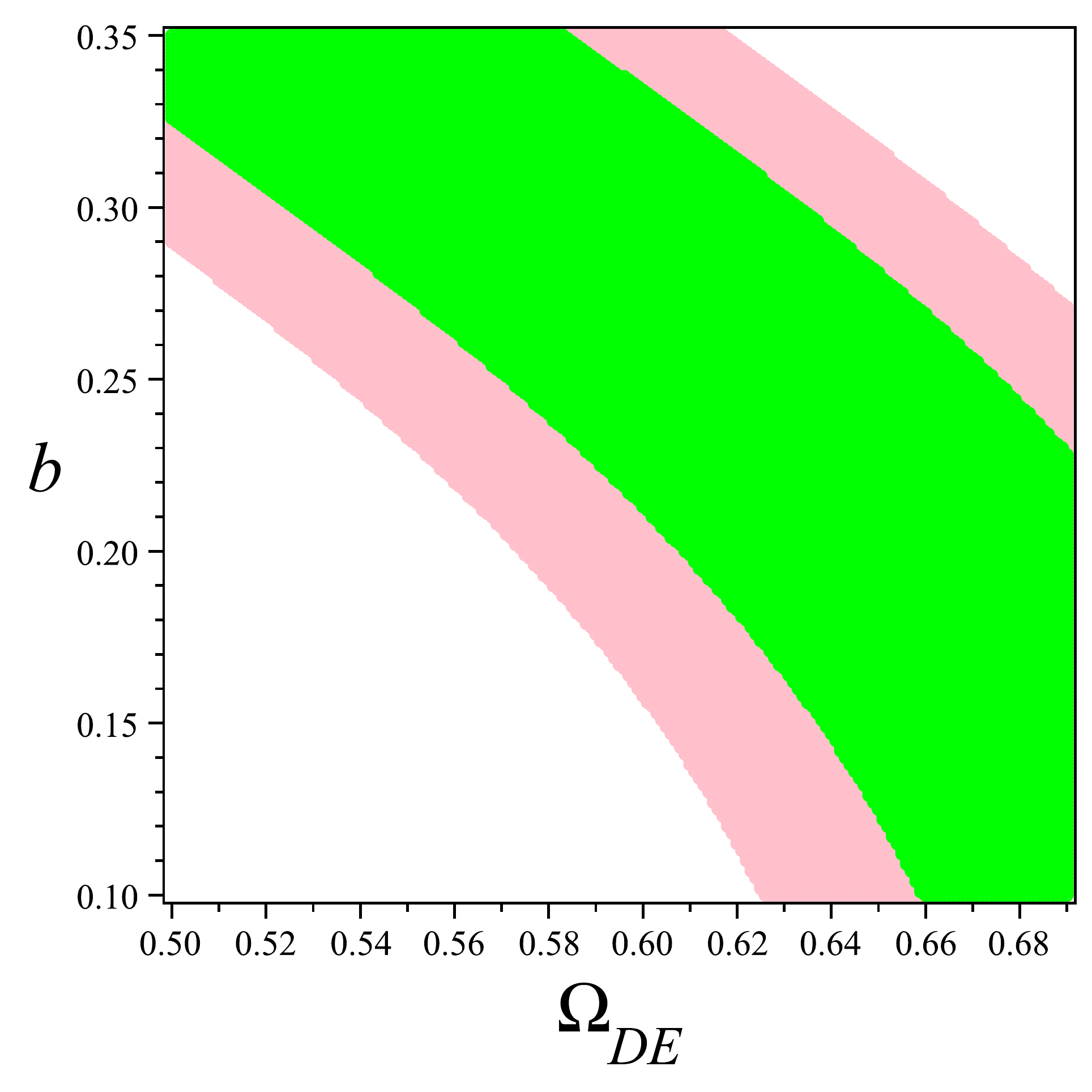}
\caption{The confidence region in 2-dim for the best fitted values of the interacting Tsallis agegraphic dark energy in the DGP braneworld cosmology }
\label{fig:8}       
\end{figure*}

\begin{figure*}
\includegraphics[width=0.43\textwidth]{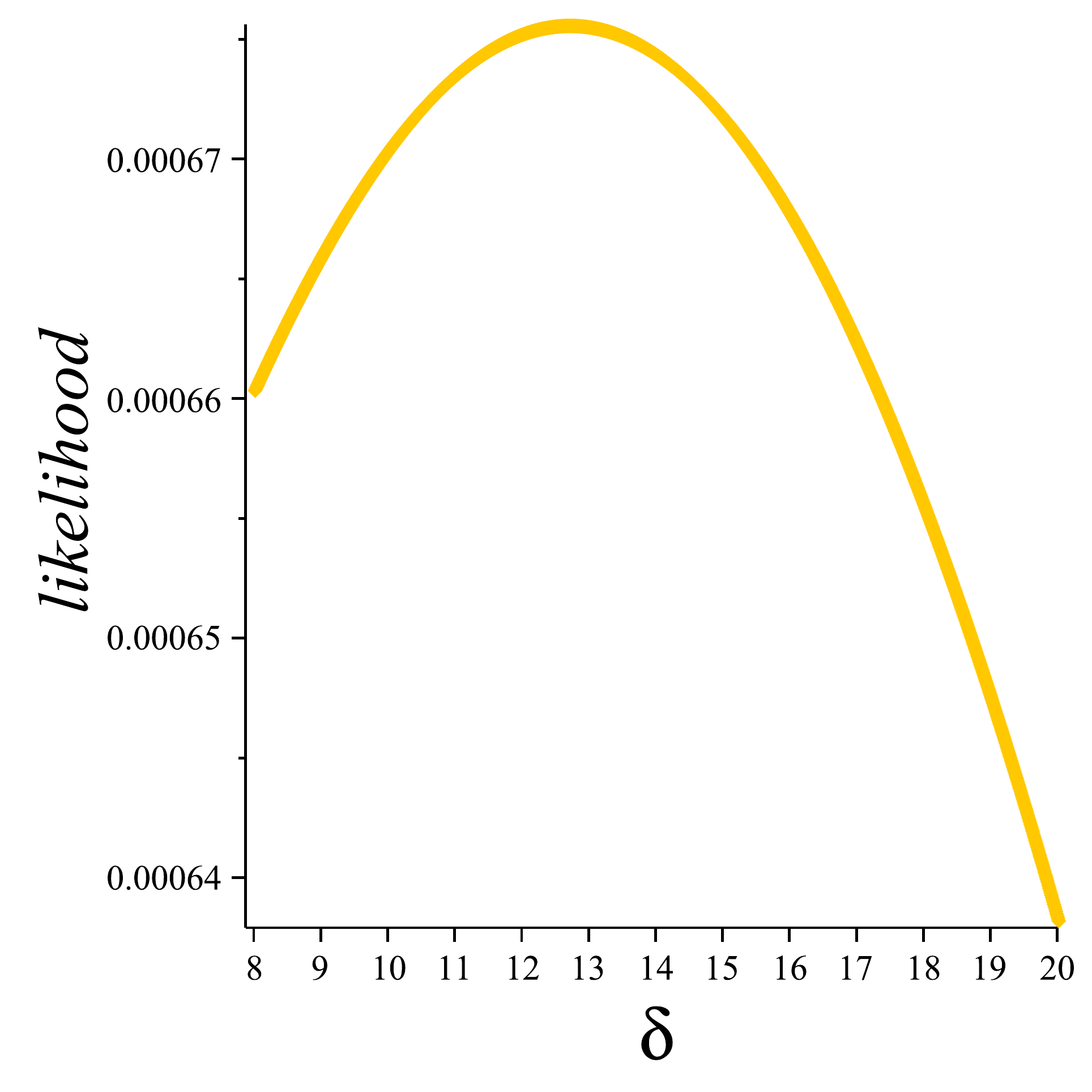}\\
\includegraphics[width=0.43\textwidth]{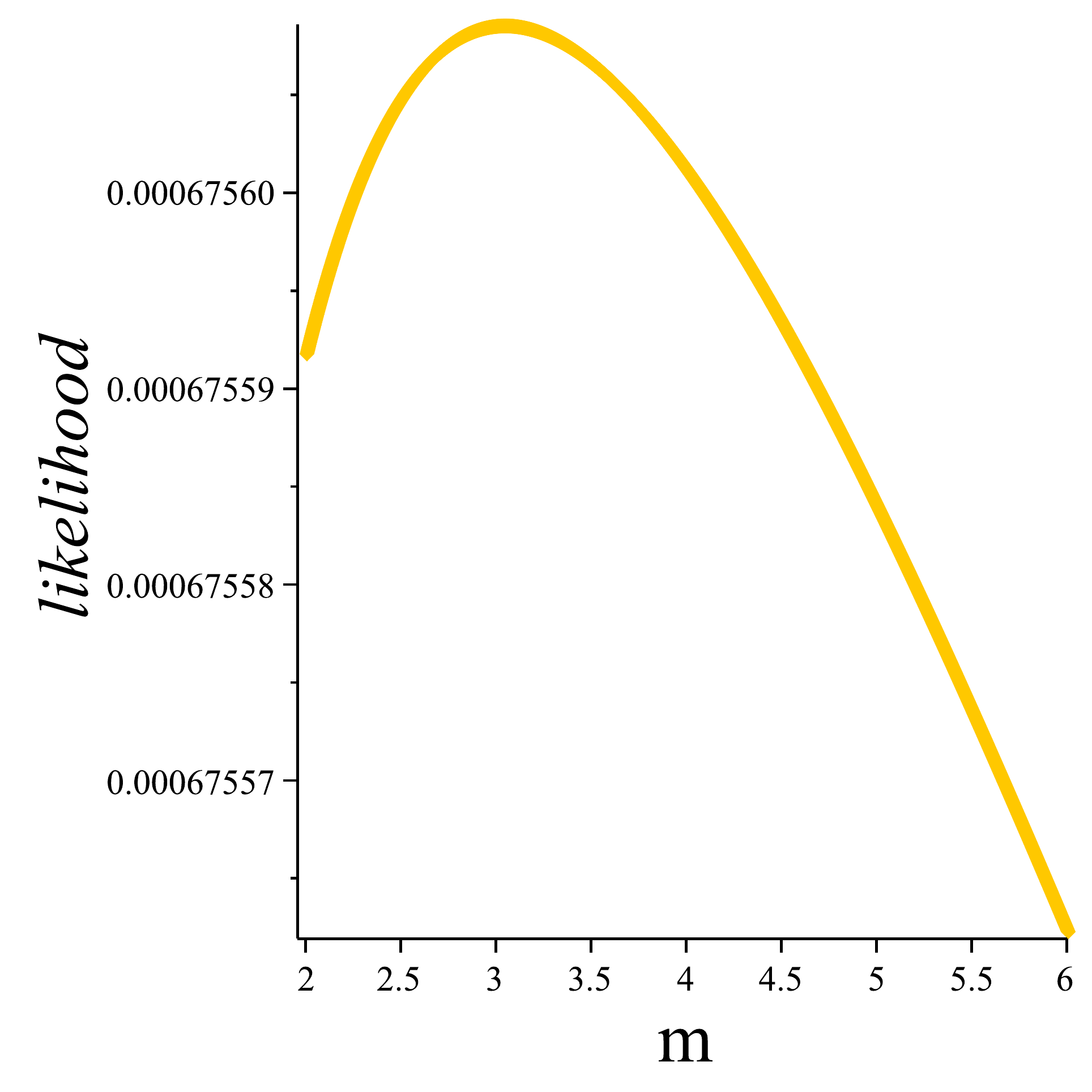}\\
\includegraphics[width=0.43\textwidth]{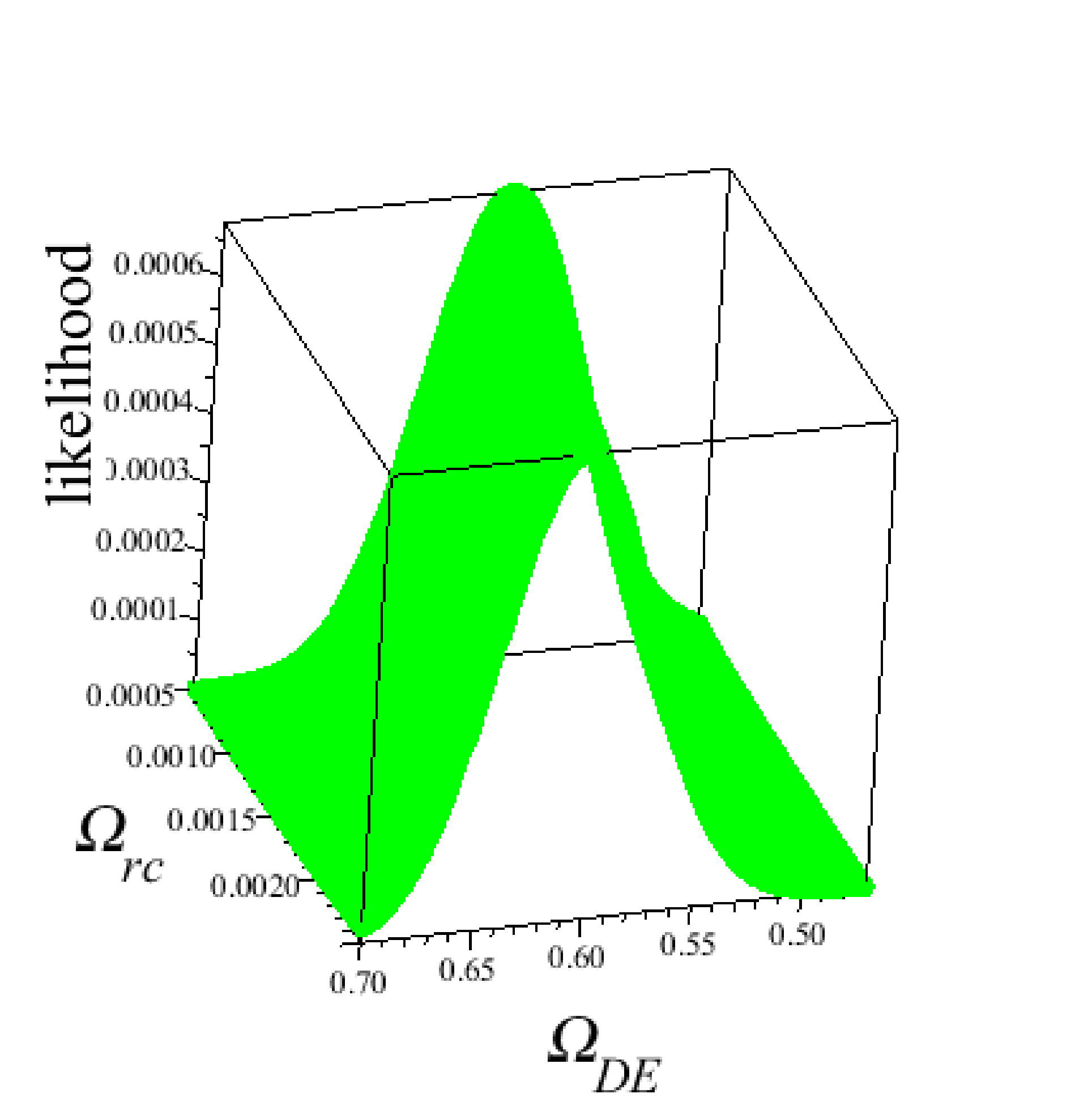}
\caption{The 1-dim and 2-dim likelihood distribution of the interacting Tsallis agegraphic dark energy in the DGP braneworld cosmology  } \label{fig:9}       
\end{figure*}

Now, we discuss the behavior of dark energy density, deceleration, dark energy equation of state, and the total equation of state parameters using the best-fit values of the model in Fig. 10 and 11. From figure 10, it is manifest that $\Omega_{DE}\longrightarrow 0$ at the early Universe ($z\longrightarrow\infty$) and $\Omega_{DE}\longrightarrow 1$ when $z\longrightarrow -0.5$. From Fig. 11 (top plot), the transition point from the decelerating to accelerating phase is $z=0.568$, and the current value of $q$ for the model with best-fit values is $q(z=0)=-0.487$. Next to that, from Fig. 11 (middle plot), we observe that the model behaves as the phantom at $z\longrightarrow\infty$ and $z\longrightarrow0$, then while it is in the phantom phase, it will approach $w_{DE}=-1$ in the future. Also, from this panel, the current value of the dark energy equation of state parameter can be identified by $w_{DE}(z=0)=-1.216$. However, the total equation of state parameter (bottom plot) with $w_{tot}(z=0)= -0.658$ shows the adding CDM contribution to the total energy density causes the model to lie in the quintessence phase in past and present. Then, the Universe will pass the cross line and go to the phantom phase in the future era, and again it tends to the $\Lambda CDM$ state during the Universe evolution.
Following the figures, in Fig. 12, we have graphed the Hubble parameter versus redshift. This graph helps to compare the model with the observational data set. Additionally, recently, \cite{bib72} have studied the HDE model (with future event horizon) for another interesting feature of the $H-z$ plane that leads to violating the Null Energy Condition (NEC) called $turning$ $point$. First, the authors have employed $\Omega_{DE}(z=0)=0.7$, $H(z=0)=100$, and $c<1$, then they have fitted the model by CMB+BAO and CMB+BAO+SNE Ia to use the best-fit parameters. The results show that the turning point exists for $c<1$, and it will be at $z\simeq -0.1$ for CMB+BAO+SNE Ia, while it will locate at $z\simeq 0.04$ for CMB+BAO data. Then, \cite{bib73} have investigated the turning point for the Barrow holographic dark energy (BHDE) model consisting of different $C$ and $\Delta$ for the Hubble horizon (H), and different $\Delta$ for the future event horizon (F). The results of the BHDEF model state that with increasing $C$ and decreasing $\Delta$, the turning point moves to the future, while for $C\geq 4.45$ and $\Delta\leq 0.04$, it will vanish. Also, they have considered that the turning point for the BHDE model using the best-fitted values of H+SNe Ia data ($C= 3.421^{+1.753}_{-1.611}$ and $\Delta= 0.094^{+0.094}_{-0.101}$) have obtained by \cite{bib74} will not exist.  Thus, the authors have mentioned that the turning point does not exist in the BHDE model. From figure 12, we see that the interacting Tsallis agegraphic dark energy model can support a turning point in the region of $-1<z<-0.5$ in the future.

\begin{figure*}
\includegraphics[width=0.43\textwidth]{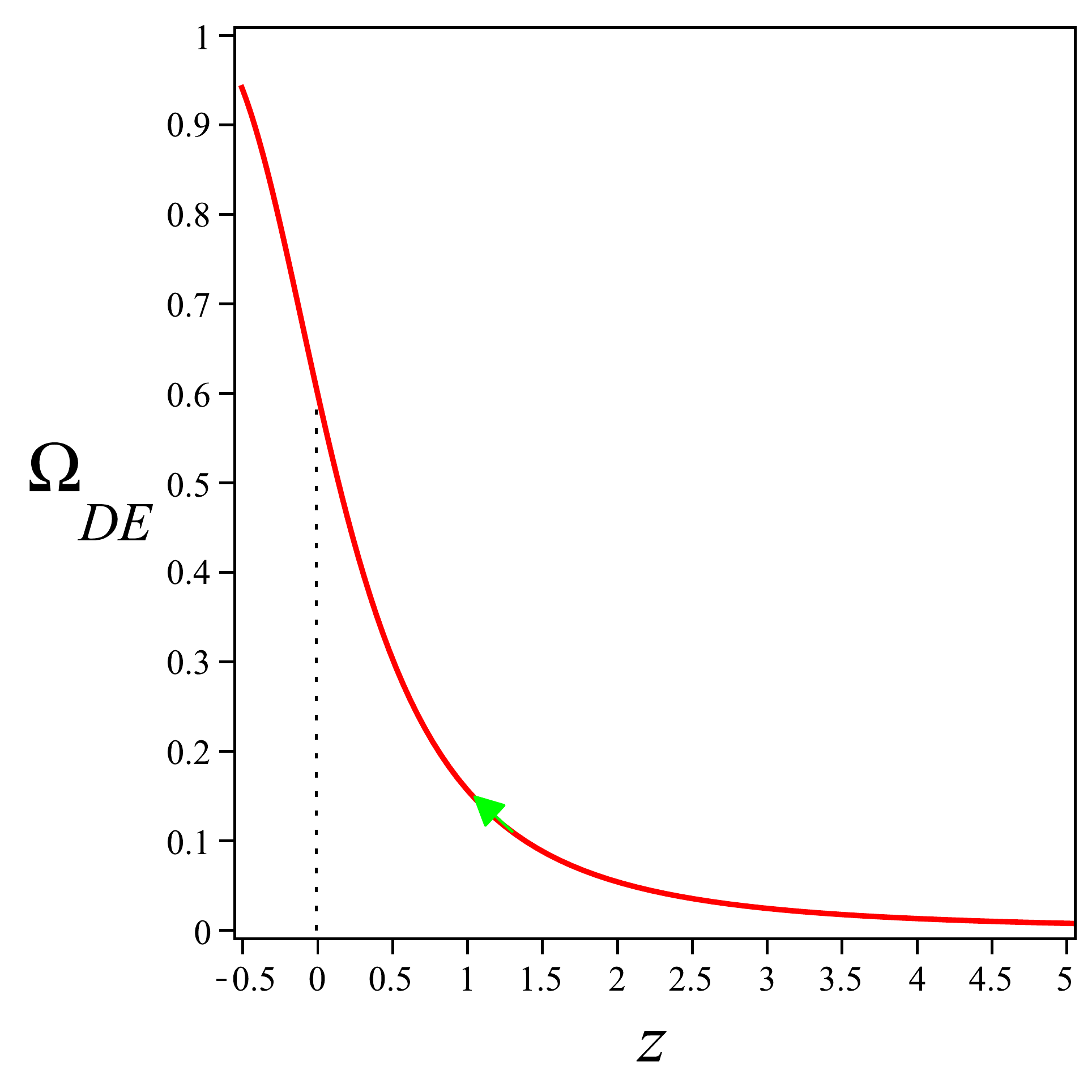}
 \caption{The evolution of $\Omega_{DE}$ versus redshift parameter z for interacting Tsallis agegraphic dark energy
 in the DGP braneworld cosmology with bestfit values  }
 \label{fig:10}       
\end{figure*}

\begin{figure*}
\includegraphics[width=0.43\textwidth]{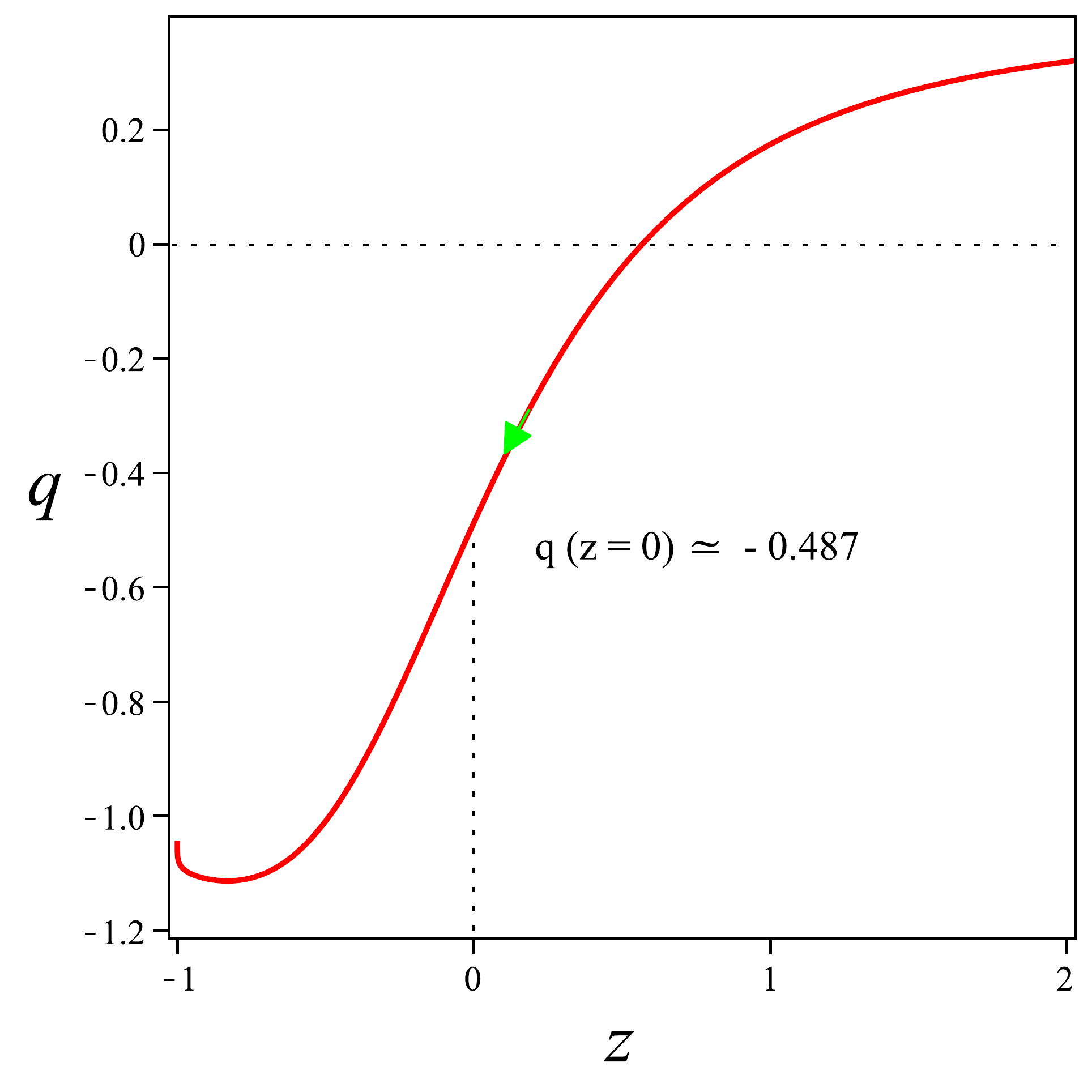}\\
 \includegraphics[width=0.43\textwidth]{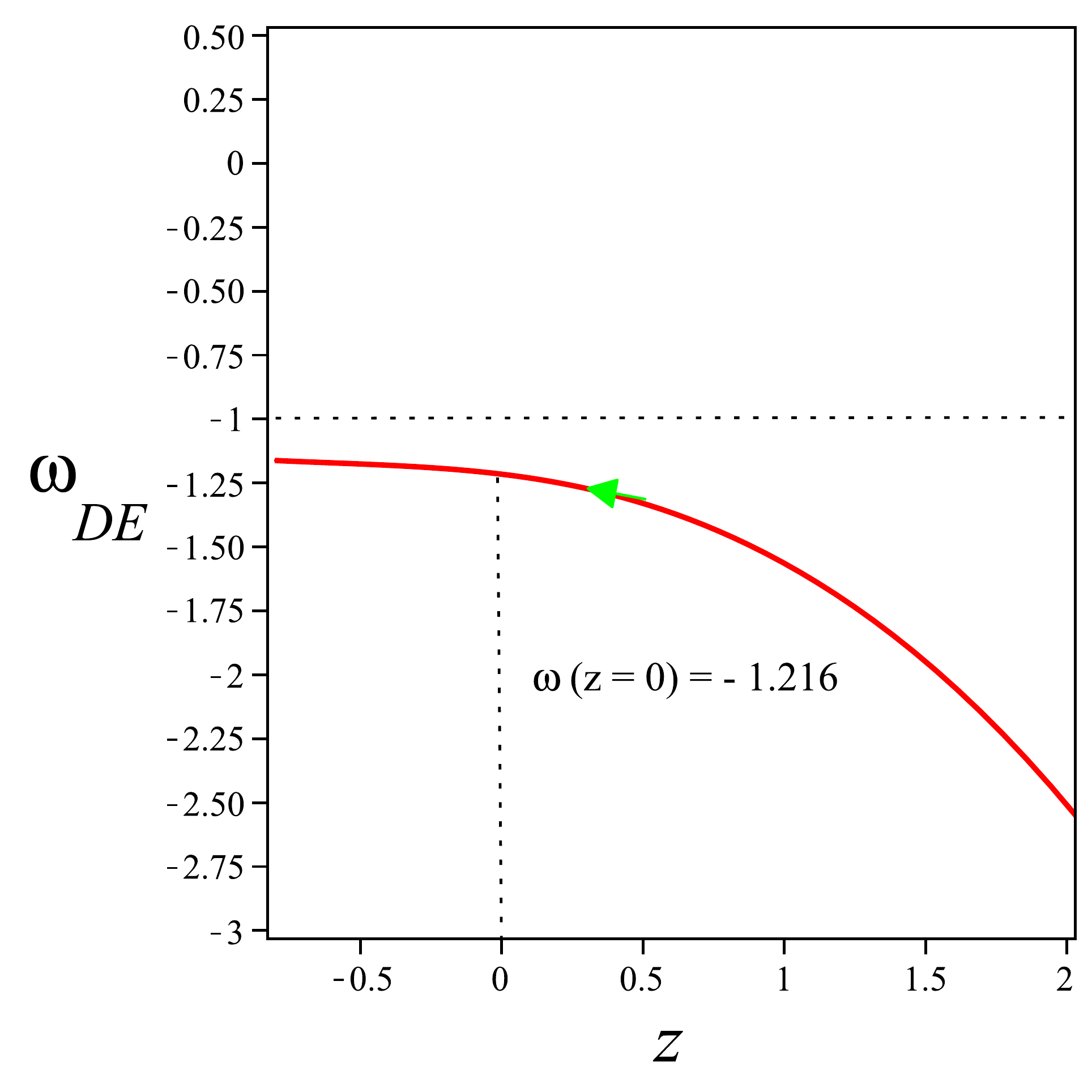}\\
 \includegraphics[width=0.43\textwidth]{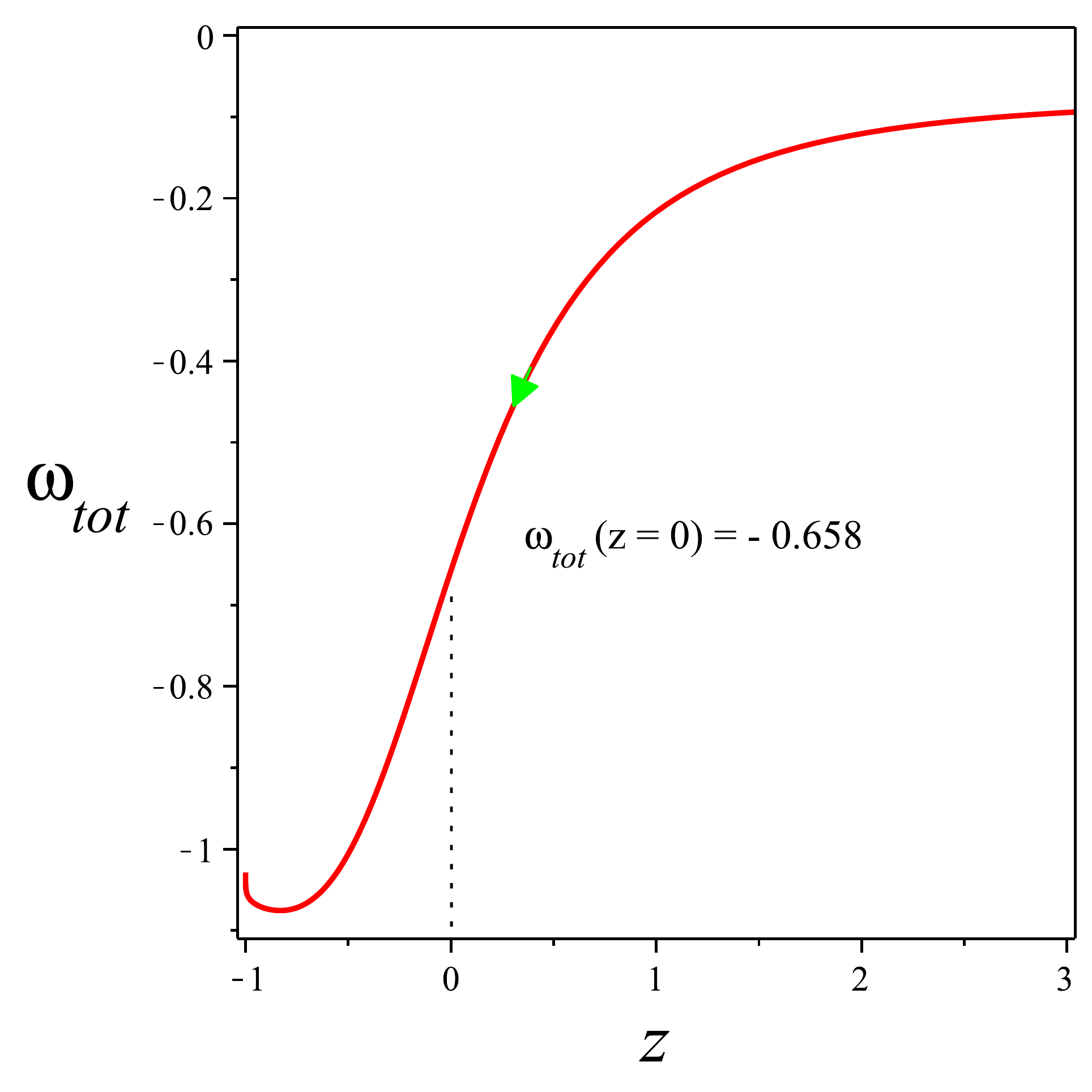}
 \caption{The evolution of q, $\omega_{DE}$ and $\omega_{tot}$ versus redshift parameter z for interacting Tsallis agegraphic dark energy
  in the DGP braneworld cosmology with bestfit values   }
 \label{fig:11}       
\end{figure*}

\begin{figure*}
\includegraphics[width=0.43\textwidth]{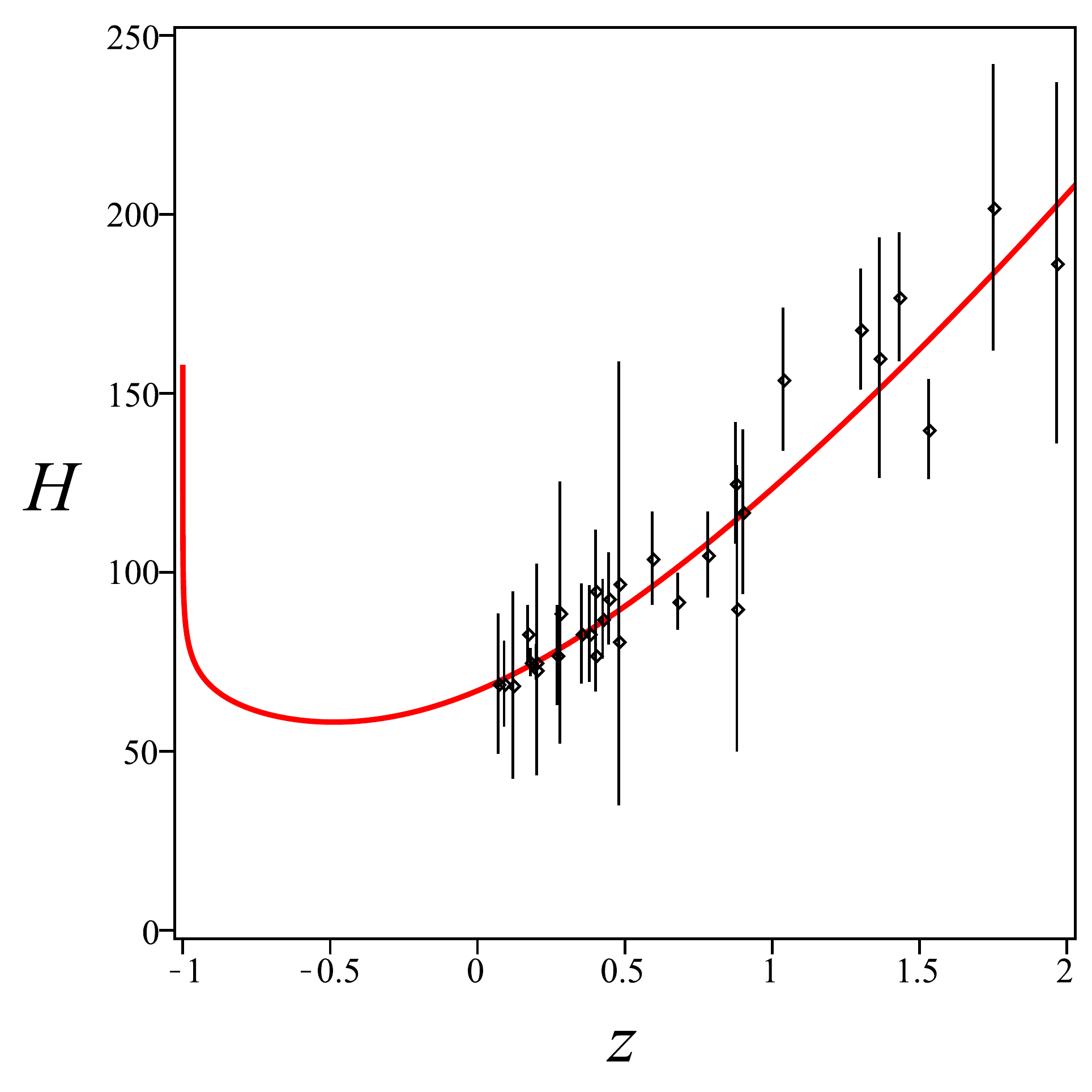}
 \caption{The Hubble parameter versus redshift in comparision with the observational data  }
 \label{fig:12}       
\end{figure*}

In Fig. 13, the $\{r,s\}$ evolutionary trajectory in the ITADE in the DGP braneworld with best-fit values is observable. The solid "Fuchsia" circle in this figure denotes the $\Lambda CDM$ fixed point ($\{r, s\}= \{1,0\}$). This figure illustrates the model behaves as the quintessence at the past and current time, then it goes toward the $\Lambda CDM$ during the cosmic evolution. The solid "Gold" circle in the panel is the present value of the statefinder parameter with $\{r_{0}, s_{0}\}\longrightarrow \{0.297, 0.237\}$.

In Fig. 14, we have plotted the  $\omega^{'}_{DE}-\omega_{DE}$ plane for the model using the best fit parameters. This plane explains that the model always remains in the thawing part during the evolution of the Universe.

\begin{figure*}
\includegraphics[width=0.43\textwidth]{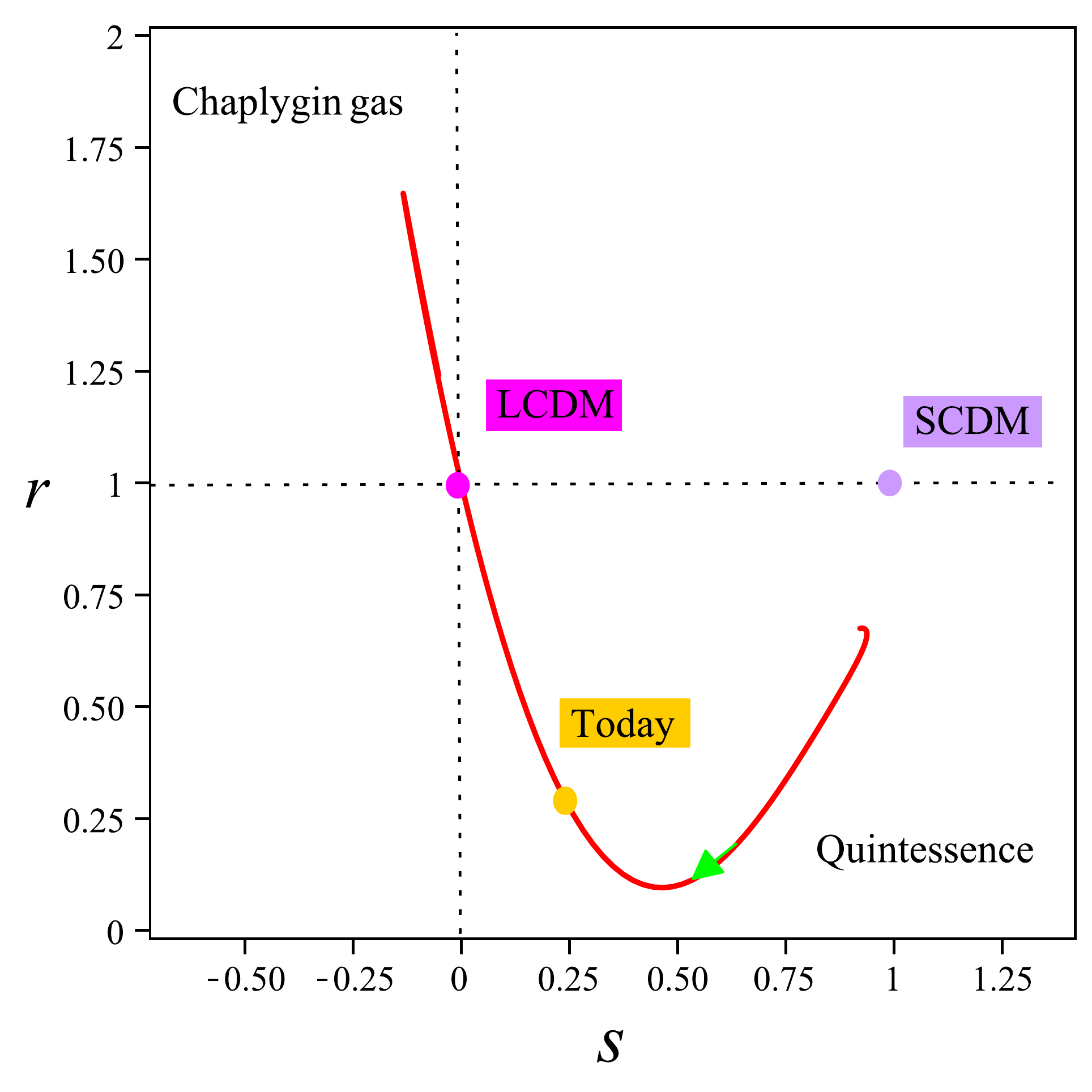}
 \caption{The evolution of r versus s for interacting Tsallis agegraphic dark energy in the DGP braneworld cosmology with bestfit values  }
 \label{fig:13}       
\end{figure*}

\begin{figure*}
\includegraphics[width=0.43\textwidth]{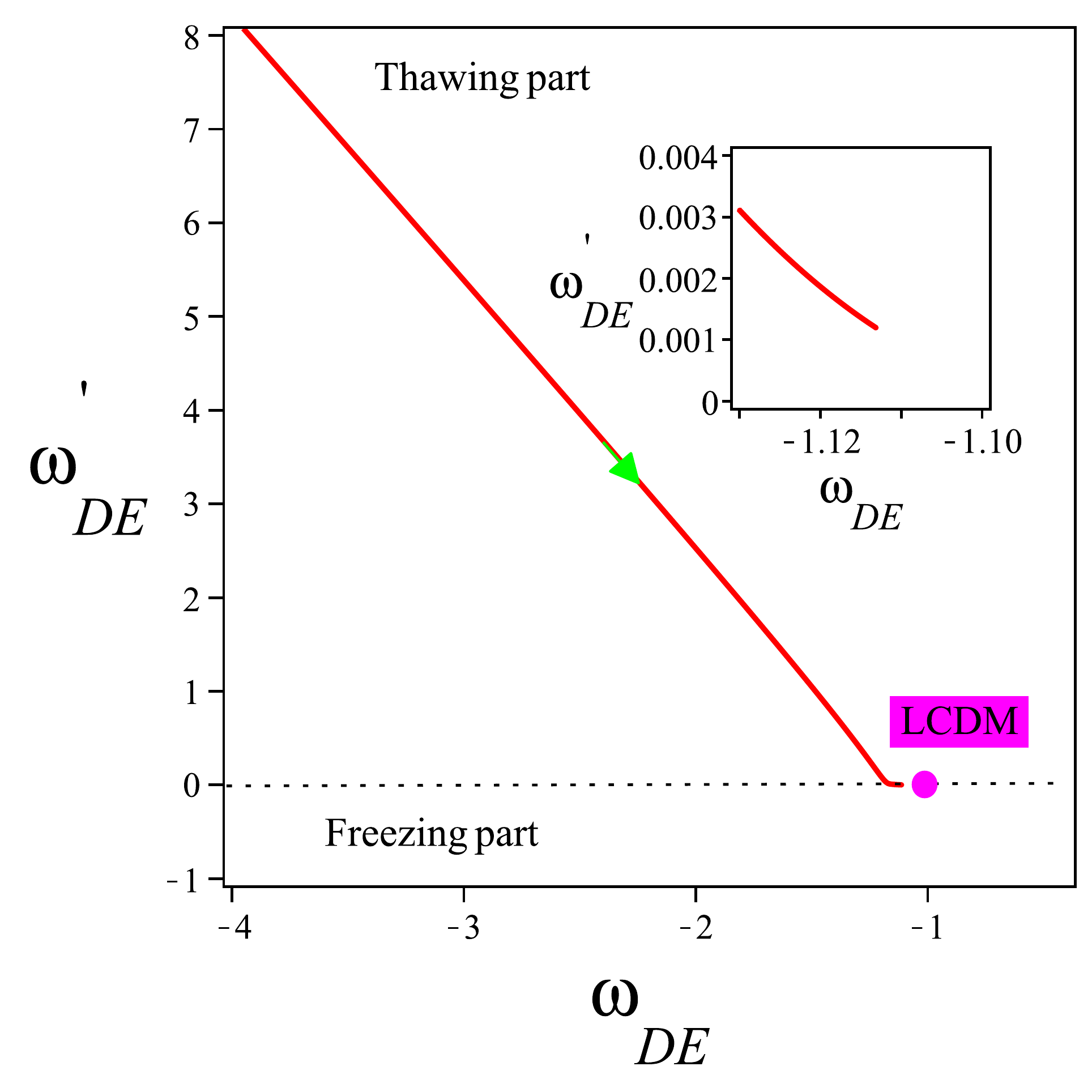}
 \caption{The $\omega^{'}_{DE}-\omega_{DE}$ diagram for interacting Tsallis agegraphic dark energy in the DGP braneworld cosmology with bestfit values  }
 \label{fig:14}       
\end{figure*}

\begin{figure*}
\includegraphics[width=0.43\textwidth]{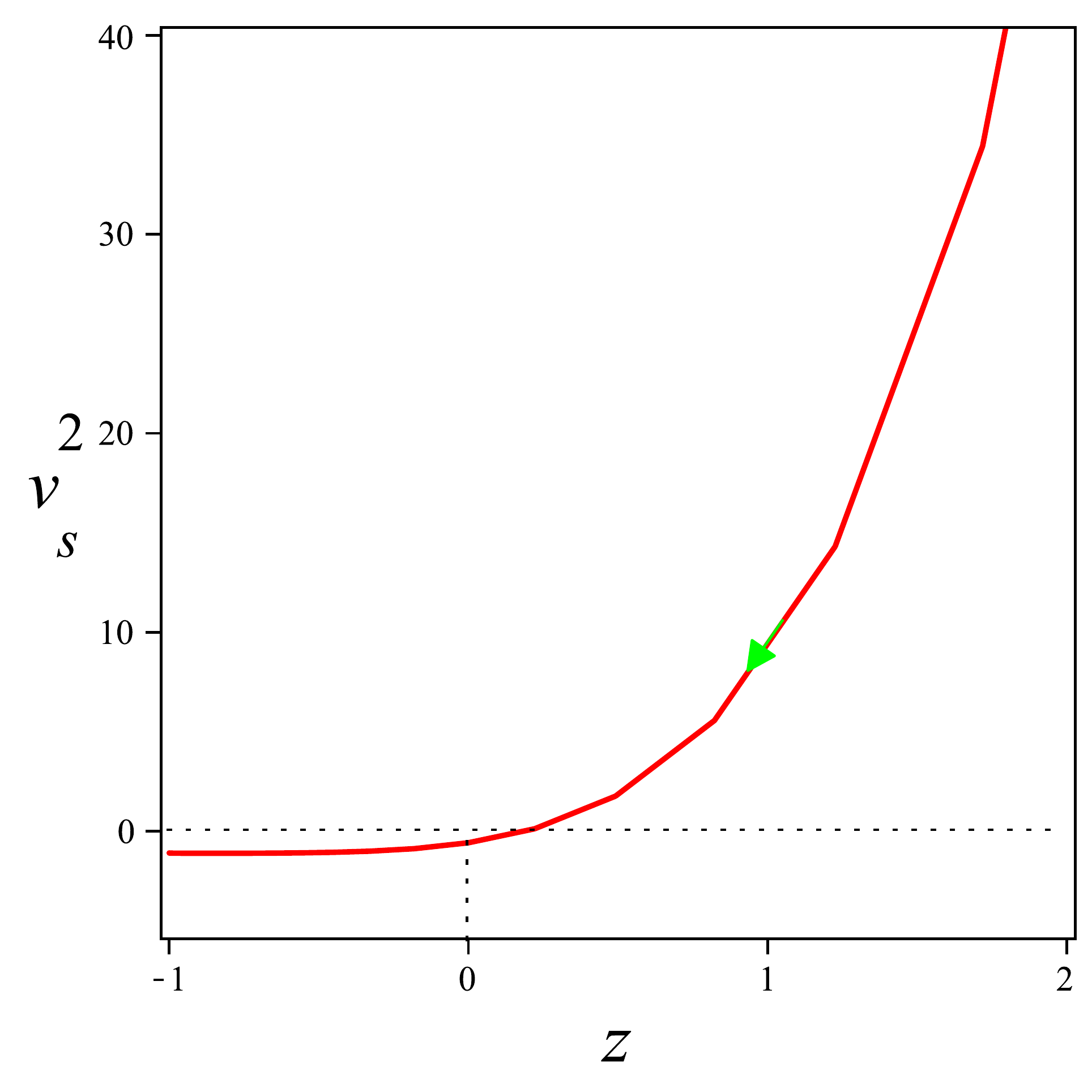}
 \caption{The evolution of $v_{s}^{2}$ versus redshift for interacting Tsallis agegraphic dark energy in the DGP braneworld cosmology with bestfit values  }
 \label{fig:15}       
\end{figure*}

Finally, the stability of the model with the best fit values is evident in Fig. 15. From this plot, the model is stable at $z\longrightarrow\infty$, while it shows an instability position at $z\longrightarrow 0$ and $z\longrightarrow -1$

\section{Conclusions}
\label{sec:7}

This paper is devoted to studying the ITADE model in the framework of the DGP braneworld scenario. First, we study the evolution of the $\Omega_{DE}$, $q$, $\omega_{DE}$, $\omega_{tot}$, $\{r-s\}$, $\omega^{'}_{DE}-\omega_{DE}$, and $v_{s}^{2}$ of the model with supposing $\Omega_{DE}(z=0)=0.680$ , $H(z=0)=66.98$, and $\Omega_{r_{c}}=0.0003$ with different $b^{2}$, $\delta$, and $m$ during the cosmic time. The deceleration parameter supports the transition from deceleration to acceleration with consistency to observational data ($0.4 < z < 1$).

Moreover, the evolution of dark energy EoS parameter against redshift for $b^{2}=0$ shows that the DE behaves as the $\Lambda$ at far past, then it lies the phantom region at the past, present, and future era. The results of different values of $b^{2}$ (except $b^{2}=0$), $\delta$, and $m$ show that the dark energy experience the phantom state in the past and current time. Furthermore, the dark energy will stay in the phantom phase while approaching -1 in the future time.

However, the total EoS parameter states that contributing the dark matter component in the total energy density causes the Universe to behave as quintessence at the past and present for all values of $b^{2}$, $\delta$, and $m$. But these values represent the Universe will enter the phantom era in the future. The results show that for $\delta=4, 5.5$, and $7$, the Universe will enter the phantom phase, then it will turn and approach the $\Lambda$ state in the future.

For further investigation, we have graphed the evolutionary trajectory of the $\{r-s\}$ pair. We have found that the model behaves as the quintessence in the past and present. Then, it will tend to the $\Lambda CDM$ in the future.

Studying the $\omega^{'}_{DE}-\omega_{DE}$ plot elucidates that for the selected values of $b^{2}$, $\delta$, and $m$, the Universe experienced the thawing feature in the past and present. In addition, the models will lie in the freezing part for all chosen values of $b^{2}$, $\delta$, and $m$, except $\delta= 4, 5.5$, and $7$, and $b^{2}= 0$. For  $\delta= 4, 5.5$, and $\delta= 7$, the Universe will stay in the thawing region, and with $b^{2}= 0$,  it goes toward the thawing part.

Furthermore, the stability analysis shows stability or instability depending on different $b^{2}$, $\delta$, and $m$. Although, the model is unstable in the future era.

Then, we have examined the ITADE in the DPG braneworld scenario with minimizing $\chi^{2}$ function with recent Hubble data. We have plotted $\Omega_{DE}$, $q$, $\omega_{DE}$, $\omega_{tot}$, $\{r-s\}$, $\omega^{'}_{DE}-\omega_{DE}$, and $v_{s}^{2}$ with best-fit parameters.

The results are approximately in agreement with results from assumed initial conditions. However, in this model with best-fit values, the $\omega^{'}_{DE}-\omega_{DE}$ is always in the thawing part.

Also, we have obtained the Akaike Information Criterion (AIC) and Bayesian Information Criterion (BIC) to compare our model with another model using the best-fitted parameters. From the AIC analysis, the model is less supported by the preferred $\Lambda CDM$, while the BIC states very strong evidence against the reference model. But, it alleviates the $\Lambda CDM$ problems. The best fit values of considered data lead to an unstable model from near the $z=0$ to the future time.


\begin{thebibliography}{9}

\bibitem{bib1} Riess, A. G., Filippenko, A. V., Challis, P., Clocchiatti, A., Diercks, A., Garnavich, P. M., Gilliland, R. L.,  Hogan, C. J., Jha, S., \& Kirshner, R. P.: Astron.~J. \textbf{116}, 1009 (1998)

\bibitem{bib2} Perlmutter, S., Aldering, G., Goldhaber, G., Knop, R. A., Nugent, P., Castro, P. G., Deustua, S., Fabbro, S., Goobar, A., \& Groom, D. E.: Astron.~J. \textbf{517}, 565 (1999)

\bibitem{bib3} Astier, P., Guy, J., Regnault, N., Pain, R., Aubourg, E., Balam, D., Basa, S., Carlberg, R. G.,
 Fabbro, S., Fouchez, D., Hook, I. M., Howell, D. A., Lafoux, H., Neill, J. D., Palanque-Delabrouille, N., Perrett, K., Pritchet, C. J., Rich, J., Sullivan, M., Taillet, R., Aldering, G., Antilogus, P., Arsenijevic, V., Balland,
  C., Baumont, S., Bronder, J., Courtois, H., Ellis, R. S., Filiol, M., Goncalves, A. C., Goobar, A., Guide, D., Hardin, D., Lusset, V., Lidman, C., McMahon, R., Mouchet, M., Mourao, A., Perlmutter, S., Ripoche, P., Tao, C.,
   \& Walton, N.: Astron. Astrophys.\textbf{447}, 31-48 (2006)

\bibitem{bib4} Riess, A. G., Kirshner, R. P., Schmidt, B. P., Jha, S., Challis, P., Garnavich, P. M., Esin, A. A., Carpenter, C., Grashius, R., \& Schild, R. E.: Astrophys.~J. \textbf{117}, 707 (1999)

\bibitem{bib5} Amanullah, R., Lidman, C., Rubin, D., Aldering, G., Astier, P., Barbary, K., Burns, M. S., Conley, A., Dawson, K. S., \& Deustua, S. E.: Astrophys.~J. \textbf{716}, 712 (2010)

\bibitem{bib6} Randall, L., \& Sundrum, R.: Phys. Rev. Lett. \textbf{83}, 3370 (1999)


\bibitem{bib7} Randall, L., \& Sundrum, R.: Phys. Rev. Lett. \textbf{83}, 4690 (1999)

\bibitem{bib8} Dvali, G. R., Gabadadze, G., \& Porrati, M.: Phys. Lett.
B \textbf{485}, 208 (2000)

\bibitem{bib9} Sahni, V., \& Starobinsky, A.A.: Int. J. Mod. Phys. D 9, 373 (2000)

\bibitem{bib10} Peebles, P.J.E., \& Ratra, B.:  Rev. Mod. Phys. \textbf{75}, 559 (2003)

\bibitem{bib11} Copeland, E. J., Sami, M., \& Tsujikawa, S.: Int. J. Mod. Phys., D \textbf{15}, 1753 (2006)

\bibitem{bib12} Armendariz-Picon, C., Mukhanov, V.F., \& Steinhardt, P.J.: Phys. Rev. Lett. \textbf{85}, 4438 (2000)

\bibitem{bib13} Kamenshchik, A.Y., Moschella, U., \& Pasquier, V.: Phys. Lett. B \textbf{511}, 265 (2001)

\bibitem{bib14} Caldwell, R.R.: Phys. Lett. B \textbf{545}, 23 (2002)

\bibitem{bib15} Cai, R. G.: Phys. Lett. B \textbf{657}, 228 (2007)

\bibitem{bib16} Wei, H. \& Cai, R. G.: Phys. Lett. B \textbf{660}, 113 (2008)

\bibitem{bib17} Wei, H.\& Cai, R.-G.: Phys. Lett. B\textbf{663}, (2008). arXiv:0708.1894

\bibitem{bib18} Liu, X.-L., Zhang, X.: Commun. Theor. Phys. \textbf{52} (2009). arXiv:0909.4911

\bibitem{bib19} Jamil, M., Saridakis, E. N.: JCAP \textbf{1007}, 028 (2010). arXiv:1003.5637

\bibitem{bib20} Saaidi, K., Sheikhahmadi, H., \& Mohammadi, A. H.: Astrophys. Space Sci. \textbf{338} (2012). arXiv:1201.0275

\bibitem{bib21} Tsallis, C., Cirto, L. J.L.: Eur. Phys. J. C \textbf{73}, 2487 (2013). arXiv:1202.2154

\bibitem{bib22} Barrow, J. D.: Phy. Lett. B\textbf{808}, 135643 (2020). https://doi
    .org/10.1016/j.physletb.2020.135643

\bibitem{bib23} Moradpour, H., Corda, C., Ziaie, A. H. \& Ghaffari, S.: EPL \textbf{127}, 60006 (2019)


\bibitem{bib24} Shababi, H. \& Ourabah, K.: Eur. Phys. J. Plus \textbf{135}, 697 (2020). https://doi.org/10.1140/epjp/s13360-020-00726-9

\bibitem{bib25} Abdollahi Zadeh, M., Sheykhi, A., \& Moradpour, H.: Modern Physics Letters A \textbf{34}, 1950086 (2019). arXiv:1810.12104

\bibitem{bib26} Srivastava, Sh., Dubey, V. C., \& Sharma, U. K.: Int. J. Mod. Phys. A \textbf{35}, 2050027 (2020)

\bibitem{bib27} Xu, Y D.: Commun. Theor. Phys. \textbf{72}, 015402 (2020). https://doi.org/10.1088/1572-9494/ab544e

\bibitem{bib28} Dvali, G., Gabadadze, G., \& Porrati, M.: Phys. Lett. B \textbf{485}, 208 (2000). arXiv:hep-th/0005016


\bibitem{bib29} Deffayet, C.: Phys. Lett. B \textbf{502}, (2001). arXiv:hep-th/0010186


\bibitem{bib30} Deffayet, C., Dvali, G.R., \& Gabadadze,G.: Phys. Rev.~D \textbf{65}, 044023 (2002). arXiv:astro-ph/0105068
\bibitem{bib31} Hu, B. \& Ling, Y.: Phys. Rev.~D \textbf{73}, 123510 (2006)

\bibitem{bib32} Chimento, L.P. \& Richarte, M.G.: Phys. Rev.~D \textbf{85}, 127301 (2012)

\bibitem{bib33} Ebrahimi, E., Golchin, H., Mehrabi, A., \& Movahed, S.M.S.: Int. J. Mod. Phys. D \textbf{26},1750124 (2017). https://doi.org/
    10.1142/\\S0218271817501243

\bibitem{bib34} Arevalo, F., Bacalhau, A. P. R., \& Zimdahl, W.: Class. Quantum Grav. \textbf{29}, 235001 (2012). arXiv:1112.5095


\bibitem{bib35} Sheykhi, A., Dehghani, M. H., \& Ghaffari, S.: Int. J. Mod. Phys. D, \textbf{25}, 1650018 (2016). arXiv:1506.02505

\bibitem{bib36} Maartens, R. \& Koyama, K.: Living Rev. Relativ. \textbf{13}, 5 (2010). https://doi.org/10.12942/lrr-2010-5



\bibitem{bib37} Amarilla, L. \& Eiroa, E. F.: Phys. Rev.~D \textbf{85}, 064019 (2012)



\bibitem{bib38} Sheykhi, A. \& Wang, B.: Mod. Phys. Lett. A \textbf{25}, 1199 (2010)


\bibitem{bib39} Sheykhi, A., Dehghani, M. H., \& Hosseini, S. E.: Phys. Lett. B \textbf{726}, 23 (2013)



\bibitem{bib40} Planck Collaboration: A\&A \textbf{641}, (2020).  https://doi.org/10.1051/0004-6361/201833910


\bibitem{bib41} Rani, S.; Azhar, N.: Universe \textbf{7}, 268 (2021). https://doi.org/10.3390/universe7080268

\bibitem{bib42} Moresco, M., Pozzetti, L., Cimatti, A., Jimenez, R., Maraston, C., Verde, L., Thomas, D., Citro, A., Tojeiro, R., \& Wilkinson, D.: JCAP \textbf{05}, 014 (2016). arXiv:1601.01701

\bibitem{bib43} Sadri, E. \& Khurshudyan, M.: Int. J. Mod. Phys. D\textbf{28}, 1950152 (2019). arXiv:1809.07595


\bibitem{bib44} Ghaffari, S., Moradpour, H., Bezerra, V. B., Morais Gra\c{c}a, J.P., \& Lobo, I.P.: Phys. Dark Universe \textbf{23}, 100246 (2019)

\bibitem{bib45} Ghaffari, S., Dehghani, M. H., \& Sheykhi, A.: Phys. Rev.~D \textbf{89}, 123009 (2014)

\bibitem{bib46} Gumjudpai, B.: Gen Relativ Gravit \textbf{36}, (2004). https://doi.org/10.1023/
    B:GERG.0000016922.72972.67

\bibitem{bib47} Jawad, A., Rani, S., Salako, I. G., \& Gulshan, F.: Eur. Phys. J. Plus \textbf{131}, 236 (2016). https://doi.org/10.1140/epjp/
    i2016-16236-x

\bibitem{bib48} Salvatelli, V., Said, N., Bruni, M., Melchiorri, A., \& Wands, D.: Phys. Rev. Lett. \textbf{113}, 181301 (2014). arXiv:1406.7297

\bibitem{bib49} Holden, D. J. \& Wands, D.: Phys. Rev.~D\textbf{61}, 043506 (2000). arXiv:gr-qc/9908026

\bibitem{bib50} Farrar, G. R. \& Peebles, P. J. E.: Astrophys.~J. \textbf{604}, 1 (2004). arXiv:astro-ph/0307316

\bibitem{bib51} Copeland, E. J., Liddle, A. R. \& Wands, D.: Phys. Rev.~D \textbf{57}, 4686 (1998). arXiv:
gr-qc/9711068

\bibitem{bib51} Copeland, E. J., Liddle, A. R. \& Wands, D.: Phys. Rev.~D \textbf{57}, 4686 (1998). arXiv: gr-qc/9711068

\bibitem{bib52} Uzan, J.-P.: Phys. Rev.~D \textbf{59}, 123510 (1999). arXiv:gr-qc/9903004

\bibitem{bib53} Amendola, L.: Phys. Rev.~D \textbf{62}, 043511 (2000). astro-ph/9908023

\bibitem{bib54} Bean, R. \& Magueijo, J.: Phys. Lett. B\textbf{517}, 177 (2001). arXiv:astro-ph/0007199

\bibitem{bib55} Bean, R.: Phys. Rev.~D \textbf{64}, 123516 (2001)
\bibitem{bib56} Das, S., Corasaniti, P. S. \& Khoury, J.: Phys. Rev.~D \textbf{73},
083509 (2006). astro-ph/0510628

\bibitem{bib57} Lee, S., Liu, G.-C. \& Ng, K.-W.: Phys. Rev.~D \textbf{73},
083516 (2006). astro-ph/0601333

\bibitem{bib58} Copeland, E. J., Sami, M. \& Tsujikawa, S.: Int. J. Mod. Phys. D\textbf{15}, 1753 (2006). arXiv:hep-th/0603057


\bibitem{bib59} Akama, A.: Lect. Notes Phys.\textbf{176}, 0001113 (1982)


\bibitem{bib60} Pavon, D. \& Zimdahl, W.: Phys. Lett. B \textbf{628}, (2005)

\bibitem{bib61} Sadeghi, J., Khurshudyan, M., Movsisyan, A., \& Farahani, H.: JCAP \textbf{12}, 031 (2013)

\bibitem{bib62} Honarvaryan, M., Sheykhi, A., \& Moradpour, H.: Int. J. Mod. Phys. D \textbf{24}, 1550048 (2015). https://doi.org/10.1142/S0218271815500480

\bibitem{bib63} Abdollahi Zadeh, M., Sheykhi, A., Moradpour, H., \& Bamba, K.: Eur. Phys. J. C \textbf{78}, 940 (2018). https://doi.org/10.1140/epjc/s10052-018-6427-3

\bibitem{bib64} Sadri, E.: Eur. Phys. J. C \textbf{79}, 762 (2019).  https://doi.org/10.1140/epjc/s10052-019-7263-9


\bibitem{bib65} Wang, B., Zang, J., Lin, C.-Y., Abdalla, E., \& Micheletti, S.: Nuclear Physics B \textbf{778}, (2007)

\bibitem{bib66} Sheykhi, A., Sadegh Movahed, M., \&Ebrahimi, E.: Astrophys. Space Sci.\textbf{339}, (2012). https ://doi.org/10.1007/s10509-012-0977-x

\bibitem{bib67} Srivastava, S., Sharma, U., \& Dubey, V.: Gen Relativ Gravit \textbf{53}, 47 (2021). https:// doi.org/10.1007/s10714-021-02818-y

\bibitem{bib68} Sahni, V., Saini, T.D., Starobinsky, A.A., \& Alam, U.: Jetp Lett.\textbf{77}, (2003). https://
    doi.org/10.1134/1.1574831

\bibitem{bib69} Caldwell, R.R. \& Linder, E.V.: Phys. Rev. Lett. \textbf{95}, 141301 (2005)

\bibitem{bib70} Akaike, H.: IEEE Transactions on Automatic Control \textbf{19}, 716-723 (1974). doi: 10.1109/TAC.1974.1100705

\bibitem{bib71} Schwarz, G.: Annals of Statistics \textbf{6}, 461-464 (1978). http://dx.doi.org/10. 1214/aos/1176344136


\bibitem{bib72} Colga\'in, E. O\', Sheikh-Jabbari,  M. M.: Class. Quantum Grav. \textbf{38}, 177001 (2021)


\bibitem{bib73} Huang, Q., Huang, H., Xu, B., Tu, F., Chen, J.: Eur. Phys. J. C \textbf{81}, 686 (2021). https:// doi.org/10.1140/epjc/s10052-021-09480-3


\bibitem{bib74} Anagnostopoulos, F. K., Basilakos, S., Saridakis, E. N.: Eur. Phys. J. C \textbf{80}, 826 (2020)

\end{thebibliography}
\end{document}